\documentclass[8.5pt,twoside]{article}
\usepackage[super,sort&compress,comma]{natbib} 
\usepackage[version=3]{mhchem}
\usepackage{times,mathptm}
\usepackage{sectsty}
\usepackage{balance} 
\usepackage[text={11.3cm,20.4cm},centering]{geometry} 
\usepackage{graphicx} 
\usepackage{lastpage}
\usepackage[format=plain,justification=raggedright,singlelinecheck=false,font=small,labelfont=bf,labelsep=space]{caption} 
\usepackage{fancyhdr}
\usepackage{amssymb}
\usepackage{amsmath}
\usepackage{subfigure}
\usepackage{multicol}

\begin{document}

\thispagestyle{plain}
\fancypagestyle{plain}{
\fancyhead[L]{\includegraphics[height=8pt]{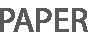}}
\fancyhead[C]{\hspace{-1cm}\includegraphics[height=15pt]{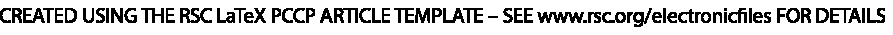}}
\fancyhead[R]{\includegraphics[height=10pt]{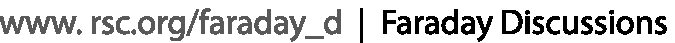}\vspace{-0.2cm}}
\renewcommand{\headrulewidth}{1pt}}
\renewcommand{\thefootnote}{\fnsymbol{footnote}}
\renewcommand\footnoterule{\vspace*{1pt}%
\hrule width 11.3cm height 0.4pt \vspace*{5pt}} 
\setcounter{secnumdepth}{5}

\makeatletter 
\renewcommand{\fnum@figure}{\textbf{Fig.~\thefigure~~}}
\def\subsubsection{\@startsection{subsubsection}{3}{10pt}{-1.25ex plus -1ex minus -.1ex}{0ex plus 0ex}{\normalsize\bf}} 
\def\paragraph{\@startsection{paragraph}{4}{10pt}{-1.25ex plus -1ex minus -.1ex}{0ex plus 0ex}{\normalsize\textit}} 
\renewcommand\@biblabel[1]{#1}            
\renewcommand\@makefntext[1]%
{\noindent\makebox[0pt][r]{\@thefnmark\,}#1}
\makeatother 
\sectionfont{\large}
\subsectionfont{\normalsize} 

\fancyfoot{}
\fancyfoot[LO,RE]{\vspace{-7pt}\includegraphics[height=8pt]{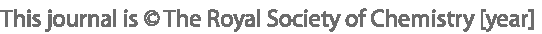}}
\fancyfoot[CO]{\vspace{-7pt}\hspace{5.9cm}\includegraphics[height=7pt]{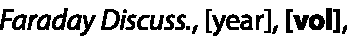}}
\fancyfoot[CE]{\vspace{-6.6pt}\hspace{-7.2cm}\includegraphics[height=7pt]{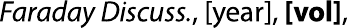}}
\fancyfoot[RO]{\scriptsize{\sffamily{1--\pageref{LastPage} ~\textbar  \hspace{2pt}\thepage}}}
\fancyfoot[LE]{\scriptsize{\sffamily{\thepage~\textbar\hspace{3.3cm} 1--\pageref{LastPage}}}}
\fancyhead{}
\renewcommand{\headrulewidth}{1pt} 
\renewcommand{\footrulewidth}{1pt}
\setlength{\arrayrulewidth}{1pt}
\setlength{\columnsep}{6.5mm}
\setlength\bibsep{1pt}


\noindent\LARGE{\textbf{Complex organic molecules along the accretion flow 
in isolated and externally irradiated protoplanetary disks}}
\vspace{0.6cm}

\noindent\large{\textbf{Catherine Walsh,$^{\ast}$\textit{$^{a}$} 
Eric Herbst,\textit{$^{b}$} 
Hideko Nomura,\textit{$^{c}$}  
T.~J.~Millar,\textit{$^{d}$} 
and 
Susanna Widicus Weaver\textit{$^{e}$}}}
\vspace{0.5cm}

\noindent\textit{\small{\textbf{Received Xth XXXXXXXXXX 20XX, Accepted Xth XXXXXXXXX 20XX\newline
First published on the web Xth XXXXXXXXXX 200X}}}

\noindent \textbf{\small{DOI: 10.1039/c000000x}}
\vspace{0.6cm}

\noindent \normalsize
{The birth environment of the Sun will have influenced the physical and chemical structure 
of the pre-solar nebula, including the attainable chemical complexity reached in the disk, important 
for prebiotic chemistry.  The formation and distribution of complex organic molecules (COMs) in a disk 
around a T~Tauri star is investigated for two 
scenarios: (i) an isolated disk, and 
(ii) a disk irradiated externally by a nearby massive star.  
The chemistry is calculated along the accretion flow from the outer disk inwards 
using a comprehensive network which includes gas-phase reactions, gas-grain 
interactions, and thermal grain-surface chemistry.  
Two simulations are performed, one beginning with complex ices and one with simple ices only.  
For the isolated disk, COMs are transported without major chemical alteration 
into the inner disk where they thermally desorb into the gas 
reaching an abundance representative of the initial assumed ice abundance.  
For simple ices, COMs can efficiently 
form on grain surfaces under the conditions in the outer disk.  
Gas-phase COMs are released into the molecular layer via photodesorption. 
For the irradiated disk, complex ices are also transported inwards; however, they undergo thermal processing caused by the 
warmer conditions in the irradiated disk which tends to reduce their abundance along the accretion flow.    
For simple ices, grain-surface chemistry cannot efficiently synthesise  
COMs in the outer disk because the necessary grain-surface radicals, which tend to be particularly volatile, are not 
sufficiently abundant on the grain surfaces.  
Gas-phase COMs are formed in the inner region of the irradiated disk via gas-phase chemistry induced by the 
desorption of strongly bound molecules such as methanol; hence, the abundances are not representative 
of the initial molecular abundances injected into the outer disk.  
These results suggest that the composition of comets formed in isolated disks may differ from those 
formed in externally irradiated disks with the latter composed of more simple ices.}
\vspace{0.5cm}

\section{Introduction}
\label{introduction}


\footnotetext{\textit{$^{a}$~Leiden Observatory, Leiden University, P.~O.~Box 9513, 2300~RA~Leiden, The Netherlands. Fax: +31~71~527~5819; Tel: +31~71~527~6287; E-mail:cwalsh@leidenuniv.nl}}
\footnotetext{\textit{$^{b}$~Departments of Chemistry, Astronomy, and Physics, University of Virginia, Charlottesville, VA~22904, USA. }}
\footnotetext{\textit{$^{c}$~Department of Earth and Planetary Sciences, Tokyo Institute of Technology, 2-12-1 Ookayama, Meguro-ku, Tokyo, 152-8551, Japan.}}
\footnotetext{\textit{$^{d}$~Astrophysics Research Centre, School of Mathematics and Physics, Queen's University Belfast, University Road, Belfast, BT7~1NN, UK.}}
\footnotetext{\textit{$^{e}$~Department of Chemistry, Emory University, Atlanta, GA~30322, US. }}



The current accepted paradigm of isolated low-mass star formation involves the 
gravitational collapse of a dense condensation within a molecular cloud forming 
a young star surrounded by a torus of dust and gas, 
the {\em protoplanetary disk}.\cite{shu87}  
The dust and gas contained within the disk provide the ingredients 
for planets and other planetary system objects such as comets.\cite{williams11}  
However, star formation rarely occurs in isolation: stars generally form in multiple 
systems and within or nearby stellar clusters.\cite{shu87,lada03}  
One of the closest and best-studied disks, TW~Hydrae, is 
associated with a small stellar cluster.\cite{kastner97}  
There is also recent evidence that the Sun may have been born within 
a stellar cluster which has long since dispersed.\cite{adams10}  
The nearest active high-mass star-forming region, the Orion molecular cloud (at a  
distance of $\approx$~400~pc), also contains many young low-mass stars which host 
protoplanetary disks, so-called {\em proplyds}.\cite{odell93}  
The proplyds in Orion are not only irradiated by the central star, they 
are also heavily irradiated and indeed photoevaporated by the external 
radiation field originating from nearby massive OB stars 
($T_\mathrm{eff}$~$\gtrsim$~10,000~K).\cite{bally98}
  
Given a large proportion of low-mass stars form 
within extreme environments, this raises questions 
regarding the influence of the external conditions on the 
physical and chemical structure of the disk.  
The birth environment may also have an impact on the planet-formation process as it 
is believed that planetesimals form via the coagulation of 
dust grains in the dense midplane.\cite{blum08}
Ice-covered dust grains can help this process: experiments have demonstrated  
that the presence of the ice mantle can significantly increase 
the sticking probability upon collision at low relative velocities.\cite{bridges96,supulver97,gundlach11} 
Ice mantles also help increase the molecular complexity 
of the gas by providing a substrate for grain-surface chemistry to build 
so-called {\em complex organic molecules} (henceforth referred to as COMs) 
which do not often have efficient gas-phase routes to formation under typical interstellar or circumstellar 
conditions.\cite{herbst09}
The chemistry of COMs in protoplanetary disks is of significance because 
these molecules may provide the `building blocks' of larger, 
more complex, {\em prebiotic} molecules, for example, amino acids.\cite{bernstein02,munoz-caro02,woon02b} 
Is it possible for such species to form within protoplanetary disks and 
survive assimilation into planets and comets?  
How does the external environment of the protoplanetary disk 
influence the resulting abundances and distribution of COMs?  
Externally irradiated disks are expected to be significantly warmer than 
isolated disks because the heating is dominated by the absorption of stellar and interstellar 
far-ultra-violet (FUV; 91.2~nm~$<$~$h\nu$~$<$~210~nm) photons by dust grains.\cite{kenyon87,chiang97,dalessio98}  
Does this warmer environment facilitate or impede the production and transport 
of COMs in the disk?  
What variation in chemical complexity is expected in the planet- and comet-forming 
zones ($\lesssim$~50~AU) of externally irradiated protoplanetary disks compared 
with isolated protoplanetary disks?

The formation of COMs in a static model of an isolated T~Tauri disk 
was investigated in previous work. \cite{walsh14}
COMs were found to form efficiently via grain-surface association reactions 
on dust grains reaching peak fractional abundances 
$\sim$~10$^{-6}$~--~10$^{-4}$ that of the H nuclei number density.  
The physical and chemical structure of 
a protoplanetary disk externally irradiated by a nearby massive star 
was also investigated. \cite{walsh13} 
Here, it was found that the disk atmosphere remains predominantly molecular in nature and 
the surface density is sufficient to effectively shield the midplane 
from the intense external FUV radiation.  
Consequently, the midplane temperature in the outer disk is 
sufficiently low ($<$~100~K) for non-volatile species ({\em e.g.}, \ce{H2O} and \ce{CH3OH}) 
to exist as ice on dust grains, potentially 
facilitating dust-grain coagulation and the formation of icy planetesimals.

In this work, the abundance and distribution of COMs 
within a typical T~Tauri disk are investigated for two local environments:  
(i) external irradiation by the interstellar radiation field only, and 
(ii) external irradiation by a nearby massive O-type star.  
The aim is to determine the influence of the external environment 
on the disk structure and the resulting implications on the formation, 
survival, and transport of COMs. 
In previous work, the chemistry evolved at each grid point
as a function of time, independent from neighbouring grid points.\cite{walsh13,walsh14} 
Here, the chemistry is calculated along streamlines which follow the 
accretion flow (and transport of material) from the outer 
disk into the planet- and comet-forming zone ($\lesssim$~50~AU).  
The influence of the accretion flow on the chemistry and 
transport of molecules within protoplanetary disks has been investigated 
previously. \cite{aikawa99,ilgner04,nomura09,heinzeller11}
However, this is the first investigation of the grain-surface 
formation and transport of COMs along the 
accretion flow in both an isolated disk and an externally irradiated 
disk.

The remainder of the paper is structured as described.  
In Section~\ref{ppdchemistry} the current observational and theoretical 
understanding of the chemical structure of protoplanetary disks 
is briefly discussed.  
In Section~\ref{ppdmodel}, the protoplanetary disk 
model is described, including the methods used to determine the 
disk physical structure (Section~\ref{physicalmodel}), the 
chemical network used (Section~\ref{chemicalmodel}), 
and the method used to calculate the chemical 
evolution along the accretion flow (Section~\ref{accretionflow}). 
In Section~\ref{results} the results of the simulations are presented and 
in Section~\ref{conclusion} the main conclusions are stated.  

\section{Chemistry in Protoplanetary Disks}
\label{ppdchemistry}

The complex interplay of radiation and physics in protoplanetary disks 
leads to disks that possess a chemically layered structure 
where the dominant chemical species reflect the dominant chemistry.\cite{bergin07}  
Photodissociation and ionisation by UV photons and X-rays 
create a surface layer abundant in radicals, atoms, and ions.  
Shielding of the UV and X-ray radiation by dust and gas 
allows ion-molecule reactions to build molecular complexity in the gas phase 
creating a warm, molecule-rich layer deeper in the atmosphere. 
Towards the disk midplane, the increasing density and decreasing temperature 
facilitate the formation (or preservation) of ice mantles on dust grains.
There is also radial stratification due to the steady increase in density and 
temperature moving inwards along the midplane towards the central star.
As the sublimation temperature of each ice species is surpassed, 
the molecule desorbs from the grain mantle and is returned to the gas phase.  
The radius at which this occurs, known as the {\em snowline}, 
depends on the molecular binding energy of each distinct ice species. 
For example, the CO snowline is expected to reside further out in the disk than the 
\ce{H2O} snowline because CO ice is more volatile than \ce{H2O} ice 
($T_\ce{CO}$~$\approx$~20~K versus $T_\ce{H2O}$~$\approx$~100~K) which 
is postulated to have implications on the C/O ratios in forming planetesimals.\cite{oberg11} 

The physical conditions in the cold, outer disk ($>$~10~AU) 
appear favourable to the production of molecules at least as complex as those observed 
in dark, molecular clouds.\cite{caselli12}
This region is best observed at far-IR to (sub)mm wavelengths 
where many molecules emit via pure rotational transitions.  
However, to date, small, simple molecules only have been detected and 
have been limited to those species which are relatively abundant and 
which possess simple rotational spectra, leading to observable line emission.   
The molecules detected in multiple disks, thus far, include CO, 
\ce{HCO+}, CN, HCN, CS, \ce{N2H+}, SO, \ce{C2H}, 
and associated $^{13}$C-, $^{18}$O-, and deuterium-containing isotopologues. 
\cite{kastner97,dutrey97,vanzadelhoff01,qi03,vandishoeck03,thi04,guilloteau06,qi08,oberg10b}
The detection of more complex species is hindered by the apparent  
size of protoplanetary disks: in astronomical terms, disks are tiny objects, 
on the order of a few arcseconds only in diameter.  
Nevertheless, current interferometric observatories have allowed 
the detection of several relatively complex molecules,
\ce{H2CO},\cite{aikawa03} \ce{HC3N},\cite{chapillon12} and {\em c-}\ce{C3H2}.\cite{qi13}
Fortunately, the Atacama Large Millimeter/Submillimeter Array (ALMA), which 
is currently in its `Early Science' phase, will have the necessary 
specifications to reach the sensitivities required to detect complex 
molecules in nearby protoplanetary disks, provided these species 
exist in the disk atmosphere with a sufficiently high abundance. 

It is now generally accepted that most COMs form within or on ice mantles on the 
surfaces of dust grains. \cite{herbst09}  
The most famous example is the simplest alcohol, methanol (\ce{CH3OH}).
Originally proposed some decades ago,\cite{allen77}
it is now known that the main route to gas-phase methanol is via the desorption 
of methanol ice which itself forms via the sequential hydrogenation 
of CO ice,
\begin{equation}
\ce{CO} \xrightarrow[(1)]{\ce{H}}\ce{HCO} 
        \xrightarrow[(2)]{\ce{H}}\ce{H2CO} 
        \xrightarrow[(3)]{\ce{H}}\genfrac{}{}{0pt}{}{\ce{CH3O}}{\ce{CH2OH}} 
        \xrightarrow[(4)]{\ce{H}}\ce{CH3OH}.
\end{equation}
Steps (2) and (4) in the above reaction scheme are assumed to possess no 
reaction barrier since they involve atom addition to a radical.   
However, steps (1) and (3) are postulated to possess large reaction 
barriers ($\approx$~2500~K), determined using quantum calculations.\cite{woon02a} 
Laboratory studies performed at 10~K have shown that steps (1) and (3) have {\em effective} barriers 
$\approx$~400~K, suggesting the reaction sequence is assisted by the 
quantum tunnelling of H atoms through the reaction barriers.\cite{watanabe02,fuchs09} 
This mechanism efficiently forms methanol ice under dark cloud conditions reaching abundances 
on the order of 1\% to 10\% that of water ice, in line with astrophysical observations.\cite{gibb04}

Modern grain-surface chemical networks now include a plethora of atom-radical and radical-radical 
reaction pathways to COMs and also precursor molecules necessary for 
forming COMs in the gas.\cite{garrod06,garrod08,laas11} 
The continuing development of these networks has been driven by the detection of 
rotational line emission from multiple COMs in so-called {\em hot cores} and {\em hot corinos}. \cite{herbst09} 
These objects are thought to be the remnant molecular material leftover from the 
process of high-mass and low-mass star formation, respectively.  
One proposed mechanism for forming COMs in these sources is that simple ices 
formed on grain surfaces at 10~K ({\em e.g.}, \ce{CO}, \ce{H2O}, \ce{NH3}, \ce{CH4}, and \ce{CH3OH}) undergo 
warming to $\approx$~20~--~30~K (caused by the ignition of the embedded star) 
where the grain-surface molecules achieve sufficient mobility for radical-radical association 
to form more complex species, {\em e.g.}, methyl formate (\ce{HCOOCH3}).\cite{garrod06,garrod08}
The necessary radicals are produced on the grain via dissociation by UV photons 
generated internally via the interaction of cosmic rays with \ce{H2} molecules.\cite{prasad83} 
The importance of radiation processing for generating complex molecules in ices has been 
known for some time, and can be efficient at low temperatures (10~K). \cite{allamandola88,gerakines96,oberg09c}
Further warming to $\gtrsim$~100~K allows those COMs formed on the grain surface to return 
to the gas phase thus `seeding' the gas with complex molecules.  
Line emission from COMs in hot cores and hot corinos is often characterised by a gas temperature 
$\gtrsim$~100~K, and abundance estimates range between $\sim$~10$^{-10}$ -- 10$^{-6}$ that of the \ce{H2} 
number density.\cite{herbst09}  

The formation of complex organic molecules in 
isolated protoplanetary disks may occur via the same mechanism 
given that a significant fraction of the disk material is $\gtrsim$~20~K.  
A fundamental difference between hot cores and protoplanetary disks is the presence 
of UV photons and X-rays which permeate the disk surface.  
In the cold, outer disk, the primary mechanism for releasing non-volatile molecules, 
such as, \ce{H2O} and \ce{CH3OH}, into the gas phase is 
desorption triggered by the absorption of UV photons, a process 
termed {\em photodesorption}. \cite{westley95} 
In the disk midplane, this process is triggered by the absorption of 
cosmic-ray-induced UV photons.
Higher in the disk atmosphere, photodesorption by stellar 
UV photons and X-rays and interstellar UV photons dominates.\cite{willacy00,walsh10}  
Photodesorption has been well-studied in the laboratory and is understood to 
be an indirect process.  
For pure CO ice, desorption from the ice surface is triggered by the 
electronic excitation of CO molecules in the bulk ice. \cite{bertin12}
For pure \ce{H2O} ice, surface desorption is induced by photodissociation of 
\ce{H2O} molecules deeper in the ice. \cite{oberg09b,arasa10}
In a protoplanetary disk, this generates a {\em photodesorbed} gas-phase layer of 
those species otherwise expected to reside on the grain at the 
temperatures in the outer disk, including \ce{H2O} and \ce{CH3OH}.  
Indeed, photodesorption is the favoured explanation for the detection 
of cold \ce{H2O} in the disk of TW~Hya.\cite{hogerheijde11}
For a protoplanetary disk externally irradiated by a nearby massive star, 
the disk material is significantly warmer ($>$~50~K); hence, 
the mechanism for complex molecule formation and release into the gas phase may 
differ from that described above for isolated systems.  

Whether COMs can form within disks also has implications on the chemical 
significance of comets, considered pristine material left over from the 
process of planet formation. \cite{mumma93}
If COMs do not form efficiently in disks, then it is possible that comets 
originally consisted of simple ices which were subsequently altered
(either via UV photons or thermal processes) 
at a more advanced phase of the evolution of the Solar System.  
Whether the Sun's natal protoplanetary disk was externally irradiated by nearby 
massive stars may also play a role in determining the attainable chemical complexity. 
A further consideration, not addressed in our previous work, 
is the composition of the material {\em entering} the protoplanetary disk.  
Is it possible for complex molecules to have formed at an earlier stage 
in the evolution of the star?  If so, how are their abundances 
altered as they are transported through the disk?
The work presented in this discussion aims to address these fundamental questions.  

\section{Protoplanetary disk model}
\label{ppdmodel}

The physical and chemical structure of a protoplanetary disk around a 
typical T~Tauri star is simulated for two scenarios: 
(i) the {\em isolated} case, in which the disk 
is irradiated by FUV and X-ray photons originating from the 
star and interstellar medium (ISM) only, 
and (ii) the {\em irradiated} case, in which the 
disk undergoes additional irradiation by FUV photons 
from a nearby ($<$~0.1~pc) massive O-type star.  

\subsection{Physical model}
\label{physicalmodel}

The physical structure of each protoplanetary disk model is 
calculated using the methods outlined in previous work.\cite{nomura05,nomura07}  
Here, the important parameters only are highlighted.  

The central star for both cases is a typical classical T~Tauri star 
with mass, $M_\ast = 0.5~\mathrm{M}_\odot$, radius, $R_\ast = 0.5~\mathrm{R}_\odot$, and 
an effective temperature, $T_\ast = 4000$~K.\cite{kenyon95} 
The disk is axisymmetric and in Keplerian rotation about the parent star. 
The physical structure of the disk is assumed to be steady ({\em i.e.}, unchanging in time) 
with a constant mass accretion rate, $\dot{M} = 10^{-8}$~$\mathrm{M}_\odot$~yr$^{-1}$.  
The kinematic viscosity in the disk, $\nu$, is parameterised using the 
dimensionless $\alpha$ parameter which scales the maximum expected size of turbulent 
eddies by the sound speed ($c_s$) and scale height ($H$) of the disk, 
$\nu \approx \alpha c_s H$.\cite{pringle81}  
For a typical T~Tauri disk, $\alpha \sim 0.01$.   

The dust temperature, gas temperature, and density structure of the disk 
are determined in a self-consistent manner by iteratively solving the 
necessary equations. \cite{nomura02}  
Initially, the physical structure of an optically thick viscous accretion disk 
in hydrostatic equilibrium is adopted. \cite{lin80} 
The specific intensity everywhere in the disk, $I_\nu(r,\phi)$, 
is calculated by solving the axisymmetric two-dimensional radiative transfer equation 
for the propagation of radiation originating from the central star and 
external sources such as the ISM and/or a nearby massive star. 
The dust temperature at each position in the disk is then recalculated assuming 
local radiative equilibrium between the absorption and reemission of radiation by 
dust grains. 
The new gas density and temperature are determined by solving the 
equations of hydrostatic equilibrium in the vertical direction 
and detailed balance between heating and cooling of the gas, respectively.  
The new disk structure is adopted and the above steps performed 
until the convergence criterion is reached. 
In this way, the FUV and X-ray mean intensities are also determined at 
each position in the disk.

Heating sources included are radiation from the 
parent star and the ISM (and the nearby massive O-type star for the irradiated disk case) 
and the radiative flux generated by viscous dissipation in the disk midplane.\cite{pringle81} 
Gas heating mechanisms included are grain photoelectric heating induced 
by FUV photons and X-ray heating due to the X-ray 
ionisation of H atoms.  The gas is allowed to cool via gas-grain 
collisions and line transitions.  

UV excess emission is commonly observed towards T~Tauri stars.\cite{herbig86} 
The radiation field from the central star is modelled as a combination of black body radiation 
at the star's effective temperature, 
optically thin hydrogenic bremsstrahlung radiation at a higher temperature 
($T_\mathrm{br}$~=~25,000~K), and Lyman-$\alpha$ line emission.  
The total FUV luminosity from the central star is $\sim$~10$^{31}$~erg~s$^{-1}$. 
For the isolated disk case, external irradiation by the interstellar radiation field 
is also included ($G_0$~$\approx$~1.6~$\times$~10$^{-3}$~erg~cm$^{-2}$~s$^{-1}$).  
The irradiated disk experiences additional irradiation by a nearby O-type star.    
The O-type star is assumed to radiate as a black body with effective temperature, 
$T_\mathrm{eff}$~=~45,000~K.  
The FUV flux at the disk surface is set to 
$G_\mathrm{ext}$~=~(4~$\times$~10$^{5}$)~$\times$~$G_0$.  
This corresponds to a distance between the disk and the massive star of $\lesssim$~0.1~pc.  
Young stars are also bright in X-rays.\cite{feigelson99} 
The X-ray emission from the central star for both cases is modelled as a fit 
to the observed {\em XMM-Newton} spectrum from the nearby classical T~Tauri star, 
TW Hya.  The total X-ray luminosity is $\sim$~10$^{30}$~erg~s$^{-1}$.
  
The main FUV opacity is absorption and scattering by dust grains.  
The dust grains are assumed to be well mixed with the gas with a 
gas-to-dust mass ratio $\sim$ 100 and   
the dust-grain size distribution adopted is that which reproduces the extinction 
curve of dense clouds.\cite{weingartner01} 
In the computation of the chemical reaction rates, for simplicity,  
a single dust-grain size and number density are used (see Section~\ref{chemicalmodel}). 
The extinction of X-rays occurs via two processes: absorption via photoionisation 
of all elements and Compton scattering by H atoms.\cite{maloney96}

\subsection{Chemical model}
\label{chemicalmodel}

The physical conditions encountered in protoplanetary disks cover a wide 
range in density, temperature, and radiation field strength.  
To compute the chemistry in all regions of the disk in an appropriate manner, 
a network which includes all possible chemical processes must be used. 
The network used is the same as in previous work \cite{walsh14} 
which originates from the expanded version of the {\em Ohio State University} 
(OSU) network. \cite{garrod06,garrod08,laas11}.  
The methods used to compute the majority of the 
reaction rate coefficients are the same as those 
outlined in previous work. \cite{walsh10}
The full chemical network consists of $\approx$~9350 reactions involving 
$\approx$~800 species.  
A brief description is given here.

\subsubsection{Gas-phase reactions.~~}

The gas-phase network includes two-body reactions, photodissociation and 
photoionisation, direct cosmic-ray ionisation, and cosmic-ray-induced 
photodissociation and photoionisation.  
The core network is expanded to include direct X-ray ionisation and 
X-ray-induced dissociation and ionisation of elements and molecules.  
This is simulated by replicating the set of cosmic-ray reactions and 
scaling the dissociation and ionisation rates by the total X-ray 
ionisation rate calculated at each grid point in the disk, $\zeta_\mathrm{xr}(x,z)$.
The photoreaction rates in the core network are normalised to the average strength of 
the unshielded interstellar radiation field, $G_0$.  

Under astrophysical conditions, many molecules dissociate primarily 
via line absorption rather than via continuum absorption.\cite{vandishoeck87,vandishoeck06}
For those species which achieve sufficiently large column densities, 
self shielding can dominate over shielding by dust grains. \cite{glassgold85}
Self shielding occurs when foreground molecular material absorbs 
the photons necessary for photodissociation to occur deeper into the atmosphere.  
The photodissociation cross sections can also overlap with those for \ce{H2}.
In this work, the self- and mutual-shielding of \ce{H2}, CO, and \ce{N2} 
are taken into account.\cite{lee96,visser09b,li13}

\subsubsection{Gas-grain interactions.~~}

The gas-phase network is supplemented with gas-grain interactions to allow 
the condensation (freezeout) and sublimation (desorption) 
of molecules onto, and from, dust grains, respectively.  
The freezeout rate onto dust grains is dependent upon 
the geometrical cross section of dust grains and the 
velocity of the gas.  
It is assumed that sticking occurs upon each collision with a dust grain, i.e., 
a sticking coefficient equal to 1 is used for all species. 
H atoms are known to stick less efficiently to grain surfaces than 
other species.\cite{matar10}
Here, the temperature dependence of the sticking coefficient 
for H atoms is taken into account.\cite{sha05,cuppen10}  

The thermal desorption rate is calculated assuming the bound molecule 
oscillates in a harmonic potential well.\cite{hasegawa92}
The thermal desorption rate depends on the binding energy of the  
molecule to the grain surface, the temperature of the dust, and the 
characteristic frequency of vibration of the molecule.  
Volatile molecules, {\em e.g.}, CO, will desorb from 
the grain mantle at lower temperatures than less volatile species, 
{\em e.g.}, \ce{H2O}.
Thermal desorption triggered by cosmic-ray heating of dust grains
is also included.\cite{hasegawa93}
This process is significant for volatile molecules only.  

For species which are strongly bound to the grain mantle, 
{\em e.g.}, \ce{H2O}, photodesorption is the primary mechanism for 
releasing molecules into the gas phase in the cold, outer disk.  
The photodesorption rate of each species depends on the flux of UV photons, 
the geometric cross section of a dust-grain binding site, the surface 
coverage of the molecule, and the molecular yield per UV photon.  
This process is now well studied in the laboratory and photodesorption 
yields are determined for \ce{CO}, \ce{N2}, \ce{CO2}, 
\ce{H2O}, and \ce{CH3OH}. \cite{oberg07,oberg09a,oberg09b, oberg09c}
For all other species, a yield of 10$^{-3}$ molecules photon$^{-1}$ is adopted, 
in line with those values already constrained in the laboratory.  
Recent experiments and simulations have also determined that photodesorption 
occurs from the top two monolayers of the ice mantle only, and 
this is accounted for in the calculation of the photodesorption rates.\cite{bertin12}  
Photodesorption triggered by all sources of UV photons are included: 
external stellar and interstellar UV photons, and UV photons generated internally
by the interaction of cosmic rays with \ce{H2} molecules.  

\subsubsection{Grain-surface reactions.~~}

A comprehensive grain-surface network is adopted to simulate 
the formation of COMs.\cite{garrod06,garrod08,laas11}  
The grain-surface association reactions are assumed to occur
via the Langmuir-Hinshelwood mechanism only, {\em i.e.}, two adsorbed 
species on the grain surface diffuse and react.  
The reaction rate coefficients are calculated using the rate equation method.\cite{hasegawa92}
This method is appropriate due to the relatively high densities,
$\gtrsim$~10$^{7}$~cm$^{-3}$, 
in the regions of the disk where grain-surface chemistry is important. 
Because of the discrete nature of dust grains, processes occurring on grain surfaces are stochastic:  
however, comparisons of stochastic models of grain-surface chemistry with 
models employing the rate equation method show that stochastic effects 
are most important at lower densities ($\lesssim$~10$^{5}$~cm$^{-2}$). \cite{stantcheva01,garrod09,vasyunin09}

Classical thermal diffusion is assumed for all species, except H and \ce{H2} which are 
also allowed to quantum tunnel through the diffusion barrier.  
This is the `optimistic' case: recent experiments investigating 
the surface mobility of H atoms on water ice have proved inconclusive regarding the 
exact contribution of quantum diffusion to the total diffusion rate.\cite{watanabe10}
The ratio of the diffusion barrier to the binding energy of each 
species is assumed to be 0.3.  
Again, this is an `optimistic' value which allows the efficient diffusion and 
reaction of relatively volatile species at $\sim$~20~K.  
For each exothermic grain-surface association reaction which leads to a single product, 
there is a probability that the product will be returned to the gas phase. 
This is the process of `reactive' or `chemical' desorption and provides an additional 
desorption mechanism for non-volatile molecules.  
A probability of 1\% is adopted in this work. \cite{garrod07}

\subsubsection{Dust-grain model.~~}
A fixed dust grain size of 10$^{-5}$~cm is adopted and the number density of 
dust grains is assumed to equal $\sim$~10$^{-12}$ that of H nuclei.  
These values correspond to so-called classical grains which possess 
$\approx$~10$^{6}$ surface binding sites per grain.  

\subsubsection{\ce{H2} formation rate.~~}

\ce{H2} molecules also form on dust-grain surfaces and the conversion 
from H to \ce{H2} and vice versa must also be taken into account.
Here, the most optimistic case is adopted: it is assumed 
that the rate of formation of gas-phase \ce{H2} equates to half the 
rate of arrival of H atoms onto the grain surfaces. 
This allows efficient formation of \ce{H2} on warm dust grains 
($\gtrsim$~25~K) in line with astrophysical observations towards 
diffuse clouds and photon-dominated regions (PDRs). \cite{gry02,habart03} 
If the explicit grain-surface route to \ce{H2} is included, 
the rate of formation of gas-phase \ce{H2} drops significantly at warm 
temperatures ($\gtrsim$~25~K) due to the volatile nature of H atoms.   
In reality, the formation rate likely lies between these two extrema.  
Several solutions have been suggested such as allowing both the 
physisorption and chemisorption of H atoms on dust grains, and 
including \ce{H2} formation via both 
the Eley-Rideal and Langmuir-Hinshelwood mechanisms. \cite{cazaux04,iqbal12,lebourlot12}
Certainly, these solutions should be explored in future protoplanetary disk models.

Note that this work does include the 
temperature-dependent behaviour of the sticking coefficient of H atoms 
which is found to decrease with increasing temperature. \cite{sha05,cuppen10}

\subsection{Accretion flow}
\label{accretionflow}

In previous work, the chemistry was computed at each point in the disk 
as a function of time, using initial abundances generated using 
a simple molecular cloud model.\cite{walsh10,walsh12,walsh13,walsh14}    
The chemistry was set up to evolve at each grid point independent of 
that in neighbouring grid points, {\em i.e.}, the rate of change in abundance of species 
$i$ at a point$(x,z)$ was given by,
\begin{equation}
\frac{\mathrm{d}n_i(x,z)}{\mathrm{d}t} = P_i - D_i \qquad \mathrm{cm}^{-3} \mathrm{s}^{-1},
\end{equation}
where $P_i$ and $D_i$ are the production and destruction rates of species, $i$, respectively.  
Here, chemical abundances are computed along the accretion flow by integrating 
the chemical evolution of a fluid parcel along streamlines from the outer edge of the 
disk inwards towards the central star.   
The rate of change in abundance of species $i$, is derived assuming,
\begin{equation}
\frac{\partial(n_i v_x)}{\partial x} = \frac{\mathrm{d}n_i}{\mathrm{d}t} = P_i - D_i \qquad \mathrm{cm}^{-3} \mathrm{s}^{-1},
\end{equation}
where $v_x$ is the radial velocity along a streamline.  
Using the continuity equation, 
\begin{equation}
\frac{\partial(n_\mathrm{T} v_x)}{\partial x} = 0 \qquad \mathrm{cm}^{-3}\, \mathrm{s}^{-1},
\end{equation}
where $n_\mathrm{T}$ is the total number density,  
the rate of change in abundance of species $i$ with radius is given by
\begin{equation}
\frac{\mathrm{d}n_i(x,z)}{\mathrm{d}x} = \frac{1}{v_x(x)}\frac{\mathrm{d}n_i(x,z)}{\mathrm{d}t} 
+ \frac{n_i(x,z)}{n_\mathrm{T}(x,z)}\frac{\mathrm{d}n_\mathrm{T}(x,z)}{{\mathrm{d}x}} \qquad \mathrm{cm}^{-2}.
\end{equation}

The radial velocity of the flow is given by
\begin{equation}
v_x(x) = - \frac{\dot{M}}{2\pi\Sigma(x)x} \qquad \mathrm{cm}\,\mathrm{s}^{-1},
\end{equation}
where $\dot{M}$ is the assumed mass accretion rate and $\Sigma(x)$ 
is the surface density of the disk at a radius, $x$.  
Streamlines, $l$, are defined as fractions of the disk scale height, 
{\em i.e.}, $l$~=~$f H$, with $0 \leq f \leq 4$.    
The disk scale height is determined at the disk midplane for all streamlines, 
\begin{equation}
H_0(x) = \frac{c_{s0}(x)}{\Omega(x)} \propto T_0^{1/2}{x}^{3/2}, 
\end{equation}    
where $c_{s0}(x) = \sqrt{k T_0(x)/m_0(x)}$, $T_0(x)$, and $m_0(x)$ 
are the sound speed, gas temperature, and mean 
molecular mass in the disk midplane, and 
$\Omega(x) = \sqrt{GM_\ast/x^3}$ is the Keplerian angular velocity at a radius, $x$.  
$G$, $M_\ast$, and $k$ represent the gravitational constant, stellar mass, 
and Boltzmann's constant, respectively.   
The disk scale height, $H$, and the geometrical height, $z$, are related 
via the assumption of hydrostatic equilibrium which defines the density structure, $\rho(x,z)$, 
\begin{equation}
\rho(x,z) = \rho_0(x)\exp{\left( \frac{-z^2}{2H(x)^2}\right )},
\end{equation}
where $\rho_0(x)$ is the density in the midplane, $z$~=~0, at a radius, $x$.   
Physical conditions, such as density, temperature, and radiation field strength, 
are extrapolated along the flow by assuming each parameter follows a power law 
behaviour ($\propto$~$\alpha x^{-\beta}$) between grid points.    

\subsection{Initial conditions}
\label{initialconditions}

The initial abundances injected into the outer edge of each streamline 
are extracted at a time of 10$^{5}$~years from a simple time-dependent 
molecular cloud model with a fixed density of 10$^{5}$~cm$^{-3}$ and a 
high visual extinction of $\gg$~10~mag.  
The chemical model used is the same as that described in Section~\ref{chemicalmodel}.  
A cosmic-ray ionisation rate of 1.3~$\times$~$10^{-17}$~s$^{-1}$ is assumed.  
The time-dependent cloud model calculation is begun 
assuming all species are in atomic form except hydrogen which is in molecular form.  
The assumed elemental abundances for H:He:C:N:O are 
1.0:9.75(-2):1.4(-4):7.5(-5):3.2(-4).  
For the heavier elements, H:Na:Mg:Si:S:Cl:Fe, the ratios are 
1.0:2.0(-9):7.0(-9):8.0(-9):8.0(-8):4.0(-9):3.0(-9).  

It is assumed that the material entering the outer region of the 
protoplanetary disk has remained shielded from the central star.    
Two sets of initial abundances are generated: one for the `cold' case 
assuming a constant gas and dust temperature of 10~K, and one for the `warm' case 
assuming a constant temperature of 30~K.  
In this way, either `pristine' dark cloud initial abundances are used, or 
it is assumed a degree of thermal processing has occurred before injection into the disk.  
The latter value of 30~K was chosen based on preceding work 
investigating the formation and distribution of COMs 
in a static model of an isolated protoplanetary disk. \cite{walsh14} 
In that work it was found that thermal grain-surface chemistry 
efficiently operates and builds COMs in the ice when the 
dust temperature exceeds $\approx$~20~--~30~K, depending on the assumed barrier 
to surface diffusion (0.3~--~0.5 times the desorption energy).  
For the lower value of the diffusion barrier, 
a peak in the abundances of grain-surface COMs was seen at a temperature
of $\approx$~30~K; hence, this temperature is adopted for the `warm' case.  
In addition, the maximum dust temperature reached at the outer edge of the isolated 
protoplanetary disk model is $\approx$~30~K.  
This higher temperature also affects the relative abundances and 
subsequent chemistry of volatile species, for example, CO.   
In the cold model, only those COMs which can form efficiently 
via atom-addition reactions on grain surfaces reach appreciable fractional abundances because 
at 10~K atoms alone have sufficient mobility to diffuse and react.  
However, at 30~K, weakly bound molecules, such as molecular radicals, 
also have sufficient mobility, allowing greater complexity to build within 
the ice mantle via radical-radical association reactions. \cite{garrod06}
Hence, for the `cold' case, the calculations begin with simple ices only, 
whereas for the `warm' case, the calculations begin with simple and complex ices. 
For the warm case, cosmic-ray-induced photodesorption releases a proportion 
of COMs from the grain mantle so that the calculations also begin with appreciable fractional abundances 
of COMs in the gas phase.   

The initial molecular fractional 
abundances (with respect to total H nuclei density) for COMs of interest in this work 
are listed in Table~\ref{table1}.  
The species highlighted here are those which 
have been observed in the gas phase in cold molecular 
clouds and/or hot cores/corinos and/or require 
methanol ice as a parent molecule.  
Also listed are the assumed binding energies for each 
species.\cite{garrod06}
Dimethyl ether, \ce{CH3OCH3}, methyl formate, \ce{HCOOCH3}, 
and glycolaldehyde, \ce{HOCH2CHO}, are all thought to form 
via the association of radicals produced either via the hydrogenation 
of CO ice, or the photodissociation of methanol ice,\cite{garrod08,oberg09c,laas11}
\begin{align}
\ce{CH3OH} + h\nu &\longrightarrow \ce{CH3}   + \ce{OH}, \\
                  &\longrightarrow \ce{CH3O}  + \ce{H},  \\
                  &\longrightarrow \ce{CH2OH} + \ce{H}. 
\end{align}
The relative branching ratios for these photodissociation 
pathways are thought to influence the resulting abundances 
of \ce{CH3CHO}, \ce{CH3OCH3}, \ce{HCOOCH3}, and \ce{HOCH2CHO}, via the surface-association reactions,
\begin{align}
\ce{CH3}   + \ce{HCO} &\longrightarrow \ce{CH3CHO}, \\
\ce{CH3O}  + \ce{CH3} &\longrightarrow \ce{CH3OCH3}, \\
\ce{CH3O}  + \ce{HCO} &\longrightarrow \ce{HCOOCH3}, \\
\ce{CH2OH} + \ce{HCO} &\longrightarrow \ce{HOCH2CHO}, 
\end{align}
respectively.
In this work, a ratio of 3:1:1 for pathways (9), (10), 
and (11) is assumed.  
This corresponds to the `standard' case in the 
study of methyl formate formation in hot cores by \citeauthor{laas11},~(\citeyear{laas11}).\cite{laas11}  

The faster diffusion rate of grain-surface radicals 
is apparent in the increased abundances of \ce{HCOOH}, \ce{CH3CHO}, 
\ce{CH3OCH3}, \ce{HCOOCH3}, and \ce{HOCH2CHO} on the 
grain surface for the warm case.  
This leads to a higher abundance in the gas: grain-surface species 
formed on the grain are released to the gas via 
non-thermal desorption (see Section~\ref{chemicalmodel}).  
Hence, the calculations for the warm case begin with 
significantly higher abundances of COMs more complex 
than methanol than for the cold case.

\begin{table}[h]
\small
\centering
\caption{Initial molecular fractional abundances and binding energies, $E_B$}
\label{table1}
\begin{tabular}{lcccc c}
\hline
              & \multicolumn{2}{c}{Cold} & \multicolumn{2}{c}{Warm} & $E_B$\\
Molecule      & Gas     & Ice     & Gas     & Ice     & (K)\\
\hline
\ce{CH3OH}    & 9.8(-10) & 1.8(-06) & 1.3(-09) & 3.0(-06) & 5530 \\
\ce{HCOOH}    & 7.0(-11) & 2.8(-11) & 1.3(-09) & 1.7(-06) & 5570 \\
\ce{CH3CHO}   & 8.7(-12) & 9.8(-11) & 7.0(-11) & 1.2(-08) & 2780 \\
\ce{CH3OCH3}  & 3.1(-14) & 1.5(-11) & 9.0(-11) & 8.1(-08) & 3680 \\
\ce{HCOOCH3}  & 1.2(-15) & 2.3(-15) & 5.6(-11) & 1.2(-07) & 4100 \\
\ce{HOCH2CHO} & 4.4(-21) & 1.8(-18) & 1.1(-11) & 6.6(-08) & 6680 \\
\hline 
\multicolumn{6}{l}{{\bf Note}: $a(b)$ = $a$~$\times$~10$^{b}$. 
Binding energies from \citeauthor{garrod08}, (\citeyear{garrod08}).} \\
\end{tabular}
\end{table}

\section{Results}
\label{results}

\subsection{Physical structure}
\label{physicalstructure}

\begin{figure}
\caption{Location and paths of streamlines along which the chemistry is calculated 
for the isolated disk (top panel) and irradiated disk (bottom panel). 
Time (shown on the top axes) increases from right to left.}
\subfigure{\includegraphics[width=\textwidth]{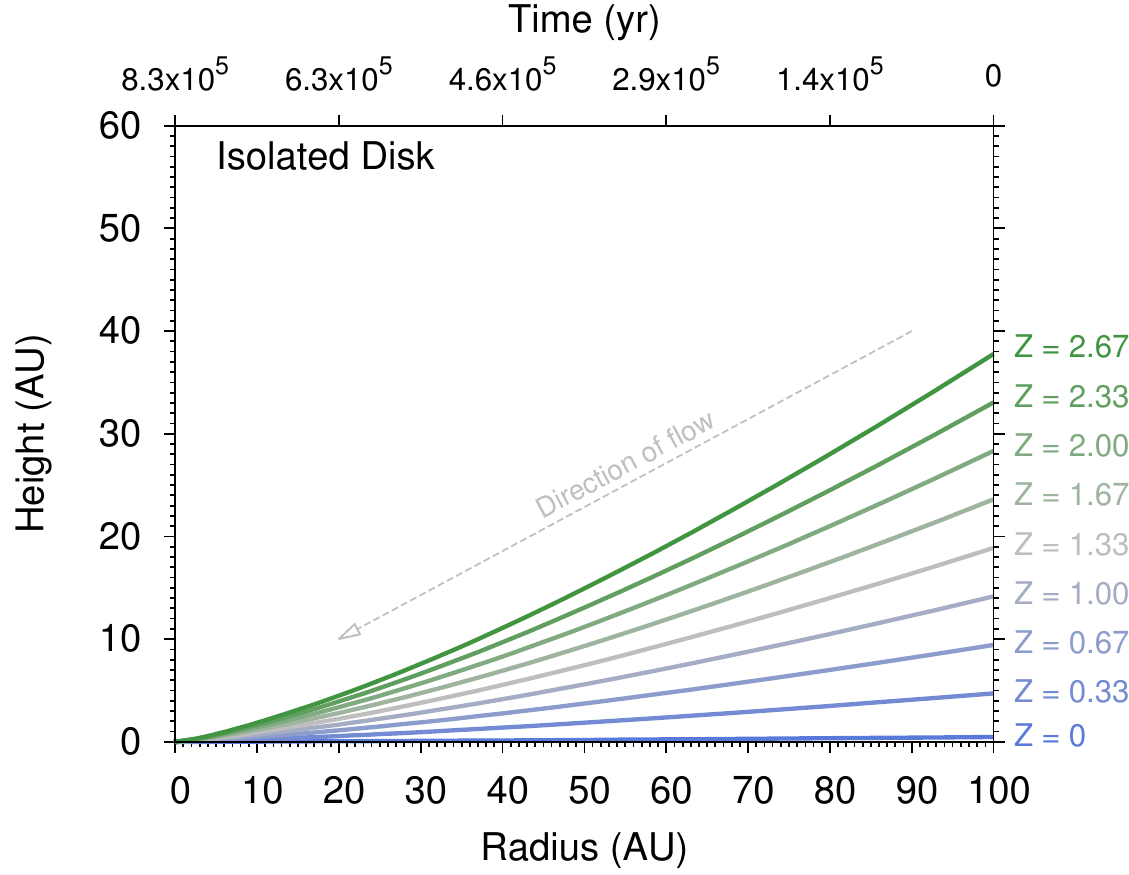}
}
\subfigure{\includegraphics[width=\textwidth]{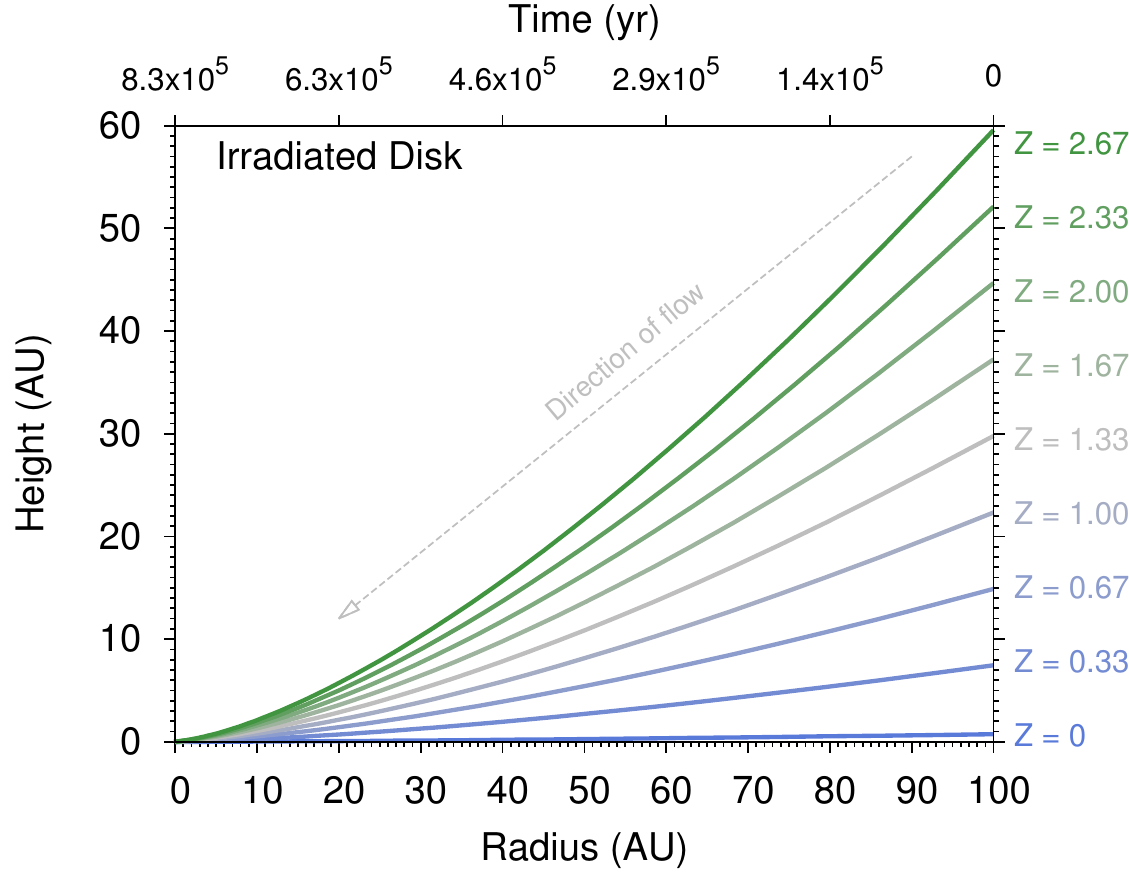}}
\label{figure1}
\end{figure}

The locations and paths of the streamlines over which the chemistry is calculated
are shown in Figure~\ref{figure1} for the isolated 
case (top panel) and the irradiated case (bottom panel).  
The chemistry evolves along each streamline from outside to inside in the direction of 
increasing time (shown in the top axes of each panel).  
The disk scale height at large radii is larger for the irradiated disk 
due to the higher gas temperature in the disk midplane.  
The streamlines considered are those which are deep enough in the disk such that 
ices remain frozen out in the outer disk.

\begin{figure}
\caption{Number density (blue lines) and gas temperature (red lines) 
along each streamline as a function of radius 
for the isolated disk (left-hand panel) and irradiated disk (right-hand panel).  
Time (shown on the top axes) increases from right to left.}
\subfigure{\includegraphics[width=0.5\textwidth]{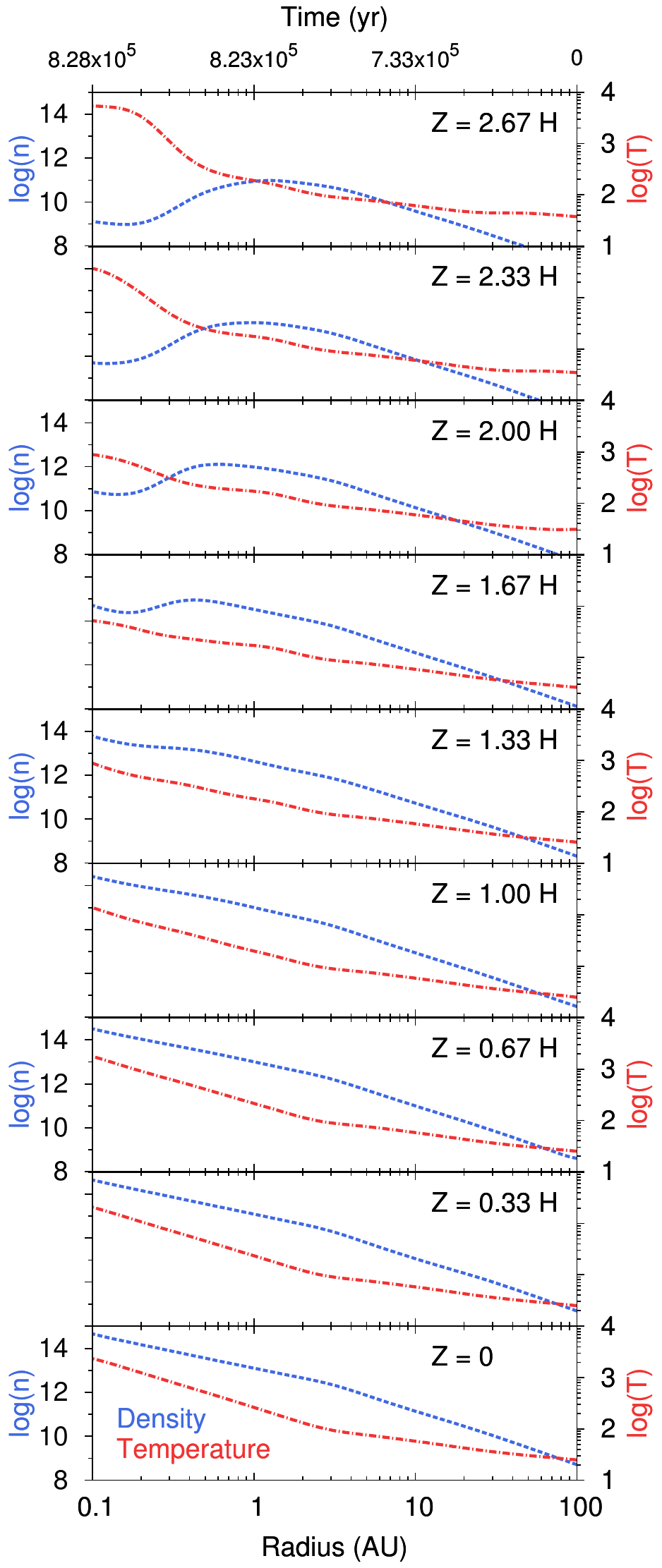}}
\subfigure{\includegraphics[width=0.5\textwidth]{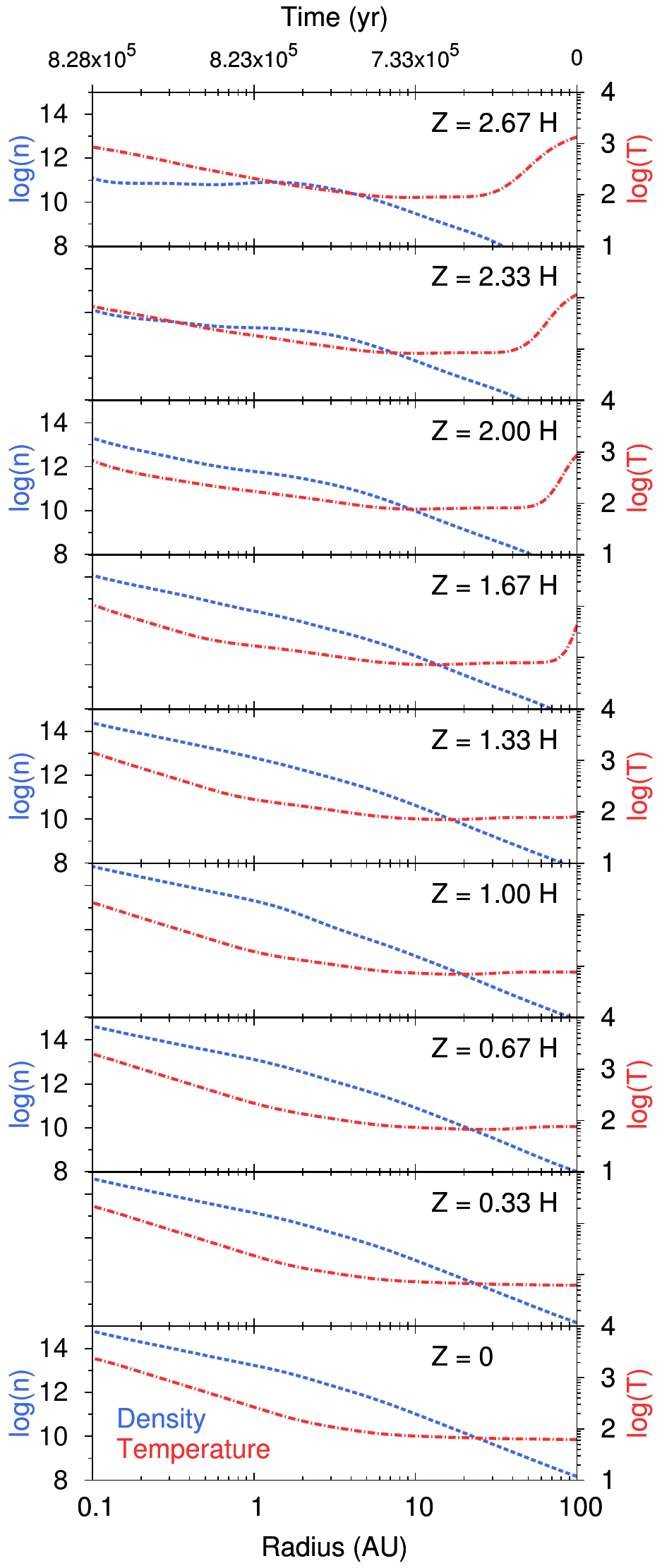}}
\label{figure2}
\end{figure}

\begin{figure}
\caption{FUV flux (green lines) and X-ray flux (gold lines) along each 
streamline as a function of radius for the isolated disk (right-hand panel) 
and irradiated disk (left-hand panel).  
The units are erg~cm$^{-2}$~s$^{-1}$.  
Time (shown on the top axes) increases from right to left.}
\subfigure{\includegraphics[width=0.5\textwidth]{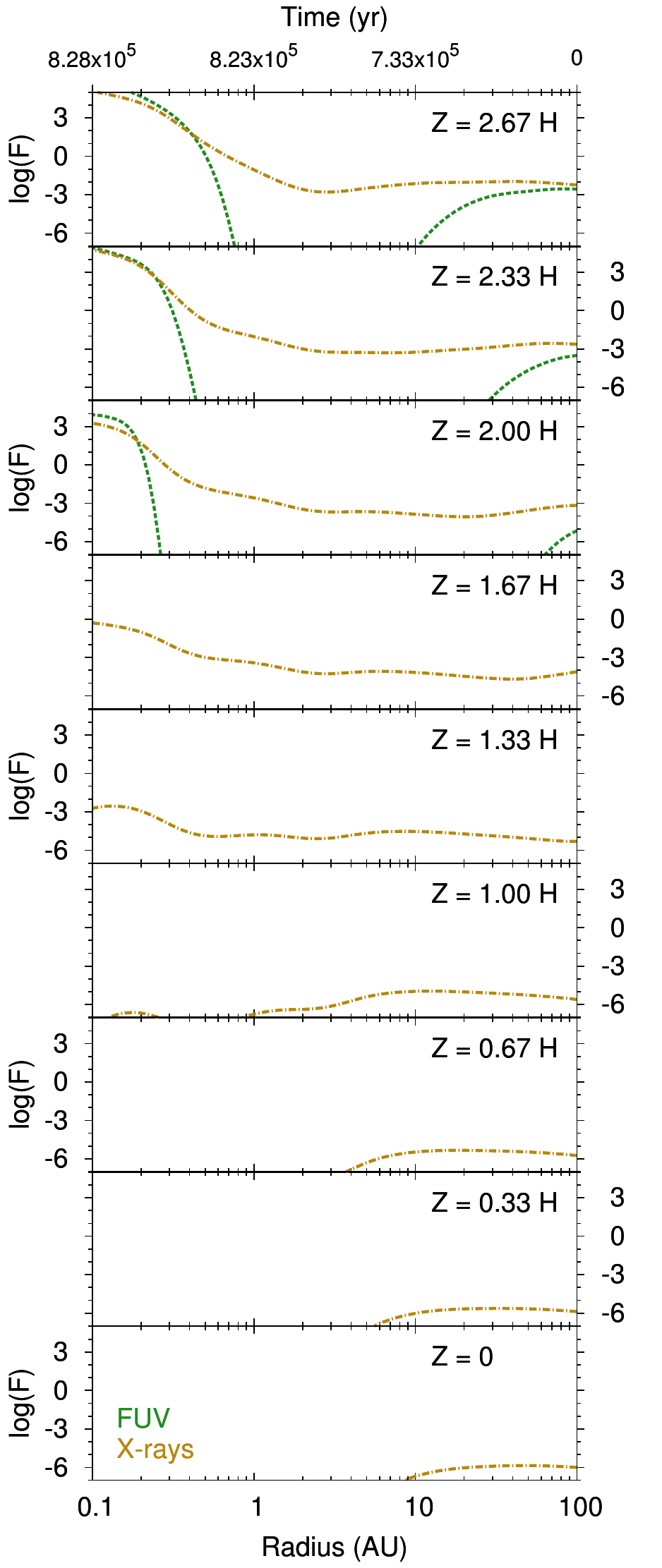}}
\subfigure{\includegraphics[width=0.5\textwidth]{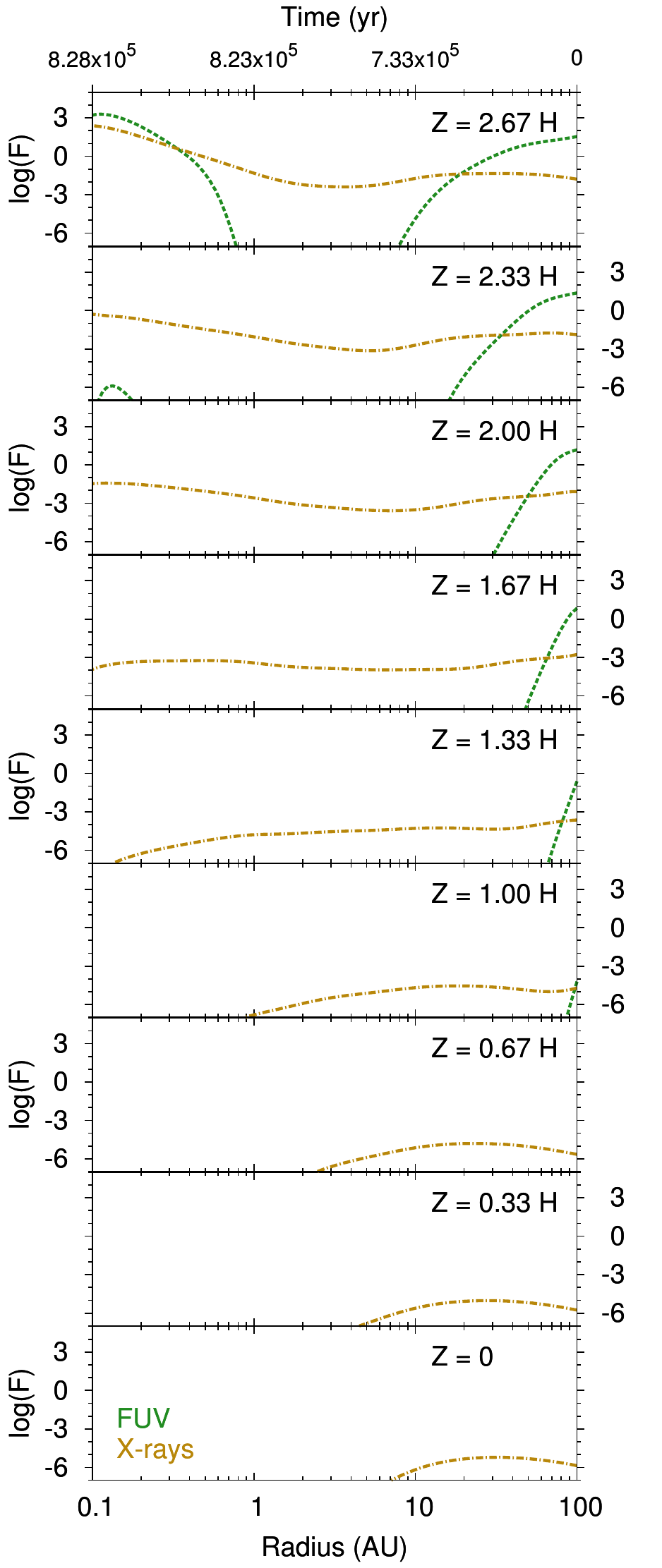}}
\label{figure3}
\end{figure}

The number density (red lines) and gas temperature (blue lines) 
along each streamline as a function of disk radius is displayed in Figure~\ref{figure2}.  
The integrated FUV flux (green lines) and X-ray flux (gold lines) 
as a function of radius are also shown in Figure~\ref{figure3}.
The data for the isolated disk are shown in the left-hand panel and those for 
the irradiated disk are shown in the right-hand panel.  

The density structure in both disk models is similar along streamlines at 
the same scale height   
because the assumed surface density is the same in both models.  
The X-ray fluxes in both models are similar for the same reason {\em i.e.}, 
the X-rays `see' a similar attenuating column of material.   
The temperature in the outer region ($\gtrsim$~10~AU) 
of the irradiated disk (right-hand panel) 
is significantly higher than for the isolated disk.  
This is due to the increased FUV flux from the nearby O-type star 
which is shown in Figure~\ref{figure3}.  
However, the temperature in the inner region ($\lesssim$~1~AU) of the 
isolated disk is higher than that of the irradiated disk for the higher streamlines 
($\gtrsim$~2~$H$).  
This is because at small radii, the disk scale height for the isolated disk 
is slightly larger than for the irradiated disk; hence, the 
higher streamlines for the isolated disk flow through regions 
of higher temperature (and thus lower density).  
This is also demonstrated by the increased FUV flux for these 
streamlines (see Figure~\ref{figure3}).  

\subsection{Chemical evolution}
\label{chemicalevolution}

The fractional abundance (with respect to H nuclei density) of each species listed 
in Table~\ref{table1} is shown in Figures~\ref{figure4} to \ref{figure13} for 
both the isolated and irradiated disk models.  
We do not show the results for \ce{CH3OCH3} because the behaviour is the same as
 that for \ce{HCOOCH3}.  
 The blue and red lines in each plot represent the results for which the 
`cold' and `warm' sets of initial molecular abundances are assumed, respectively.     
There are several general trends.  
The ice mantle survives to larger scale heights in the isolated disk 
than in the irradiated disk ($Z$~$\lesssim$~2~$H$ versus $Z$~$\lesssim$~0.67~$H$).
The stronger UV field in the outer regions of the irradiated disk efficiently strips the 
ice from the grain surface via photodesorption.  
Gas-phase COMs are subsequently destroyed by photodissociation.    
The results suggest gas-phase COMs do not survive in the molecular layers 
of externally irradiated disks.  
The gas-phase COMs in both models reach their highest fractional abundance 
in the inner disk midplane.  
The origin of the gas-phase COMs is either thermal desorption from the ice 
mantle or gas-phase formation.  
The dominant process depends on the composition of the ice mantle 
entering the inner disk.  
The initial abundances ({\em i.e.}, cold or warm) are not as critical for the 
isolated disk as for the irradiated disk.  
For those COMs which begin with a negligible abundance on 
the grain for the cold case, these species are formed efficiently 
on the grain surface under the conditions in the outer isolated disk 
(see, {\em e.g.}, Figures~\ref{figure8}, \ref{figure10}, and \ref{figure12}). 
On the other hand, for the irradiated disk, the initial composition is critical.  

The specific behaviour and chemistry of the species predicted 
by reactions (12) to (15) are now discussed.   

\subsubsection{Methanol.~~} 

\begin{figure}
\caption{Fractional abundance (with respect to number density) 
of \ce{CH3OH} gas and ice along each streamline as a function of 
disk radius for the isolated disk model. 
The blue lines show results from the model using the `cold' (10~K) set of initial abundances 
and the red lines from the model using the `warm' (30~K) set of initial abundances (see, Table~\ref{table1}).}
\includegraphics[width=\textwidth]{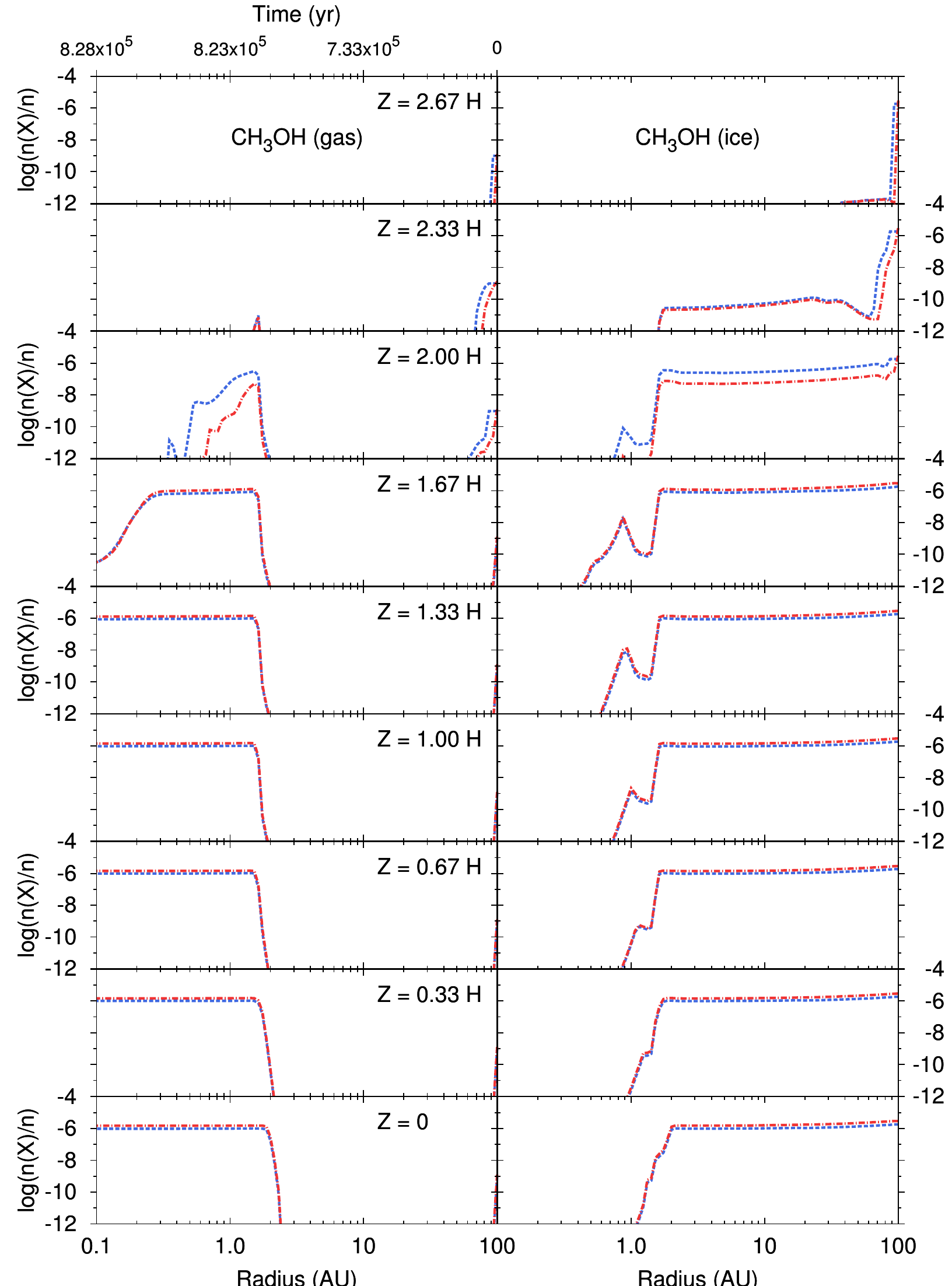}
\label{figure4}
\end{figure}

\begin{figure}
\caption{Fractional abundance (with respect to number density) 
of \ce{CH3OH} gas and ice along each streamline as a function of 
disk radius for the irradiated disk model.
The blue lines show results from the model using the `cold' (10~K) set of initial abundances 
and the red lines from the model using the `warm' (30~K) set of initial abundances (see, Table~\ref{table1}).}
\includegraphics[width=\textwidth]{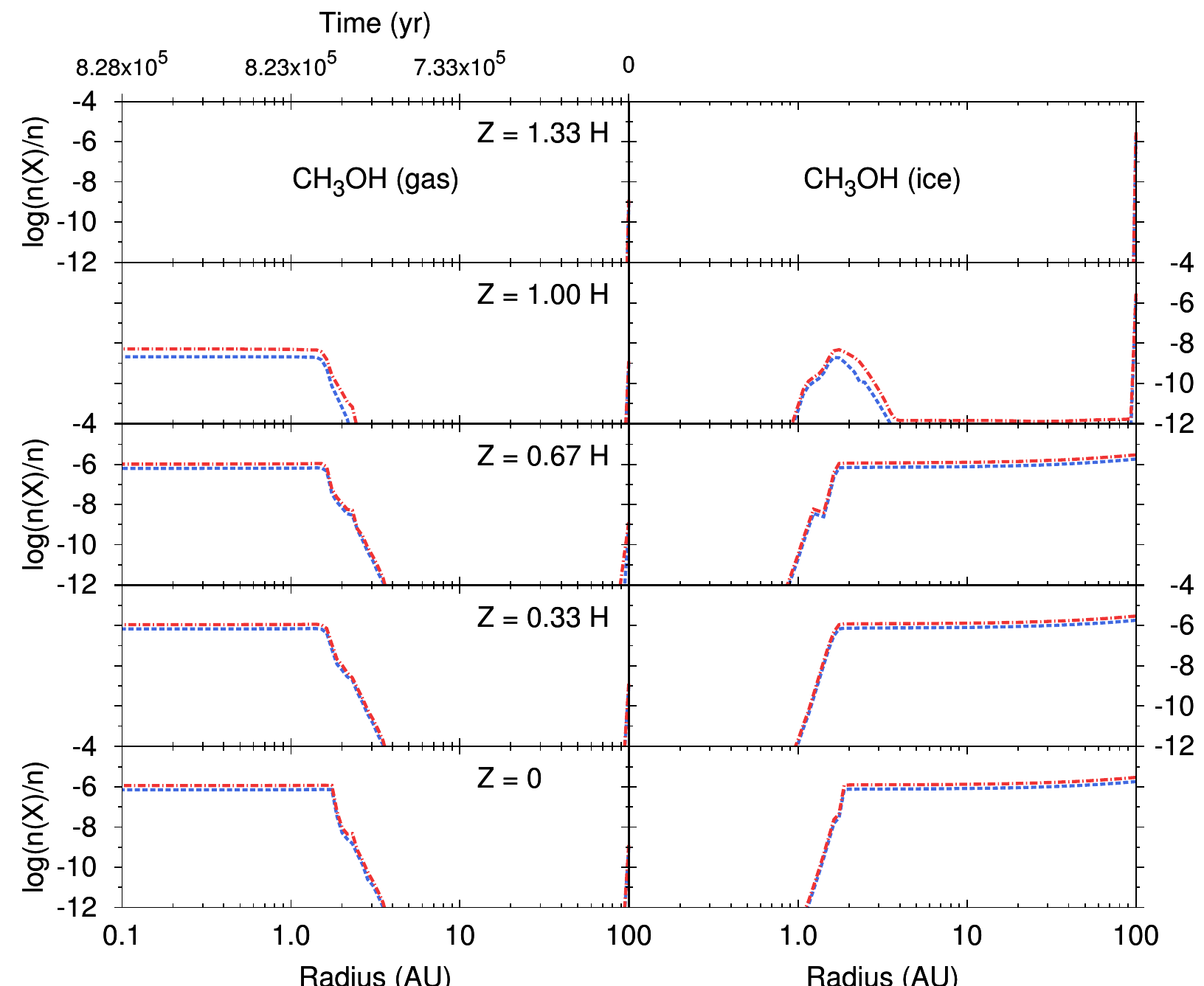}
\label{figure5}
\end{figure}

In Figures~\ref{figure4} and \ref{figure5} 
the fractional abundance of gas-phase (left-hand panel) and grain-surface (right-hand panel) 
methanol, \ce{CH3OH}, is shown as a function of disk radius for the isolated and irradiated disk models, 
respectively.  

The methanol abundance is preserved along the accretion flow in the midplane in both models.  
At $\approx$~2~AU, the temperature is sufficiently high for methanol 
ice to thermally desorb from the grain surface into the gas phase.  
This transition region is known as the {\em snowline} and resides at the same 
radius for both models.
This is because the heating in the inner disk is dominated by 
the central star and viscous dissipation rather than external irradiation.  
The abundance reached in the gas phase is the same as that for 
the ice injected into both models.   
This suggests that the methanol entering the planet- and comet-forming region 
in both irradiated and isolated protoplanetary disks has an interstellar origin.  

At higher scale heights in the isolated disk, 
methanol is photodesorbed into the gas phase reaching a fractional abundance 
$\sim$~10$^{-9}$ with respect to number density at radii $\gtrsim$~50~AU.  
This photodesorbed layer is not present in the irradiated case.  
Whether gas-phase molecules survive in the photodesorbed layer requires a delicate balance 
between photodesorption and photodissociation in the molecular layer of the disk.   
The results suggest that gas-phase methanol may be observable in isolated protoplanetary 
disks as was found in the static model.\cite{walsh14}

There are only minor differences between the cold and warm sets of abundances 
for both disk models because methanol ice reaches similar fractional abundances 
in the cold and warm cloud models ($\approx$~10$^{-6}$).

\subsubsection{Formic acid.~~} 

\begin{figure}
\caption{Fractional abundance (with respect to number density) 
of \ce{HCOOH} gas and ice along each streamline as a function of 
disk radius for the isolated disk model.
The blue lines show results from the model using the `cold' (10~K) set of initial abundances 
and the red lines from the model using the `warm' (30~K) set of initial abundances (see, Table~\ref{table1}).}
\includegraphics[width=\textwidth]{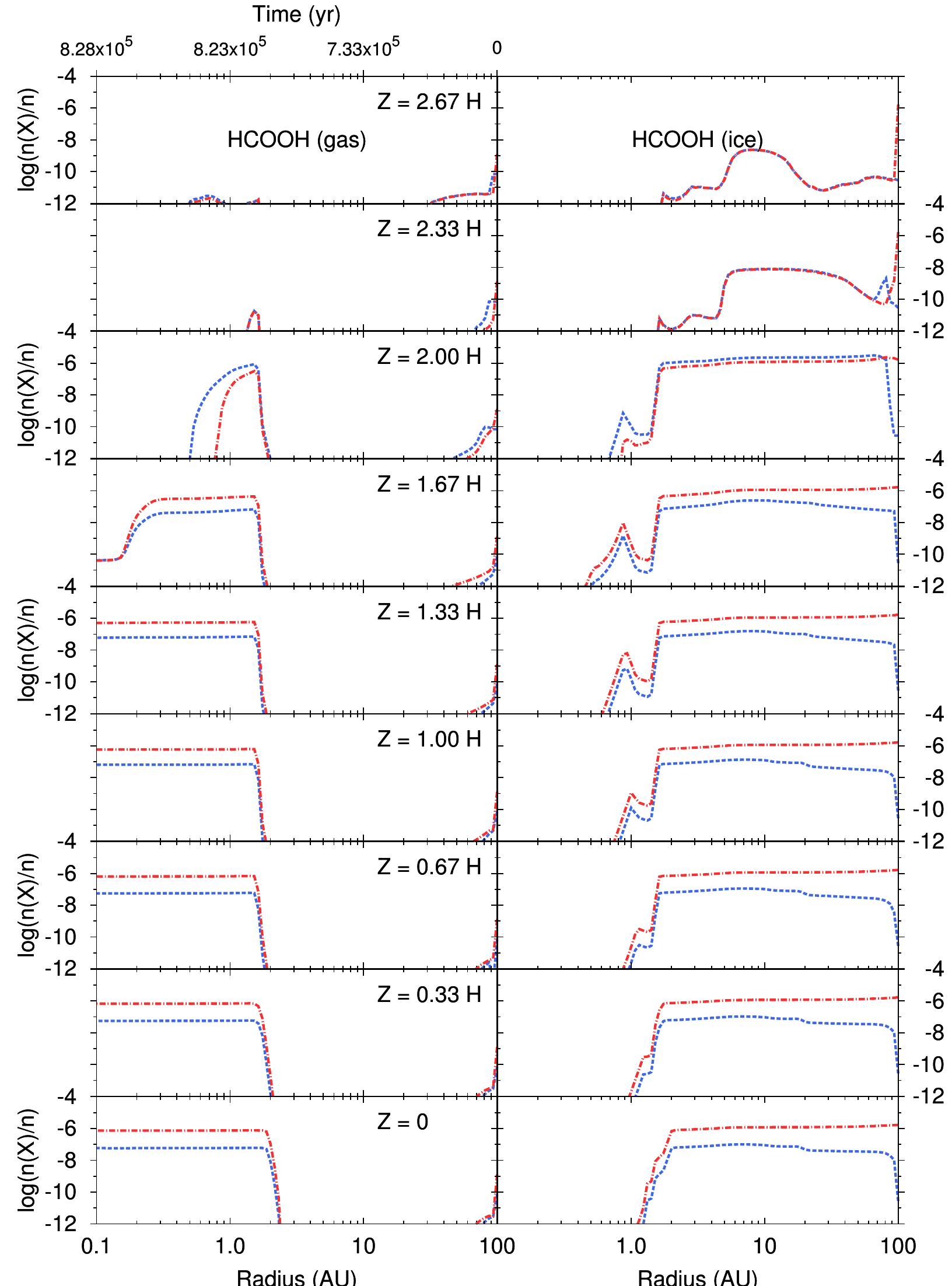}
\label{figure6}
\end{figure}

\begin{figure}
\caption{Fractional abundance (with respect to number density) 
of \ce{HCOOH} gas and ice along each streamline as a function of 
disk radius for the irradiated disk model. 
The blue lines show results from the model using the `cold' (10~K) set of initial abundances 
and the red lines from the model using the `warm' (30~K) set of initial abundances (see, Table~\ref{table1}).}
\includegraphics[width=\textwidth]{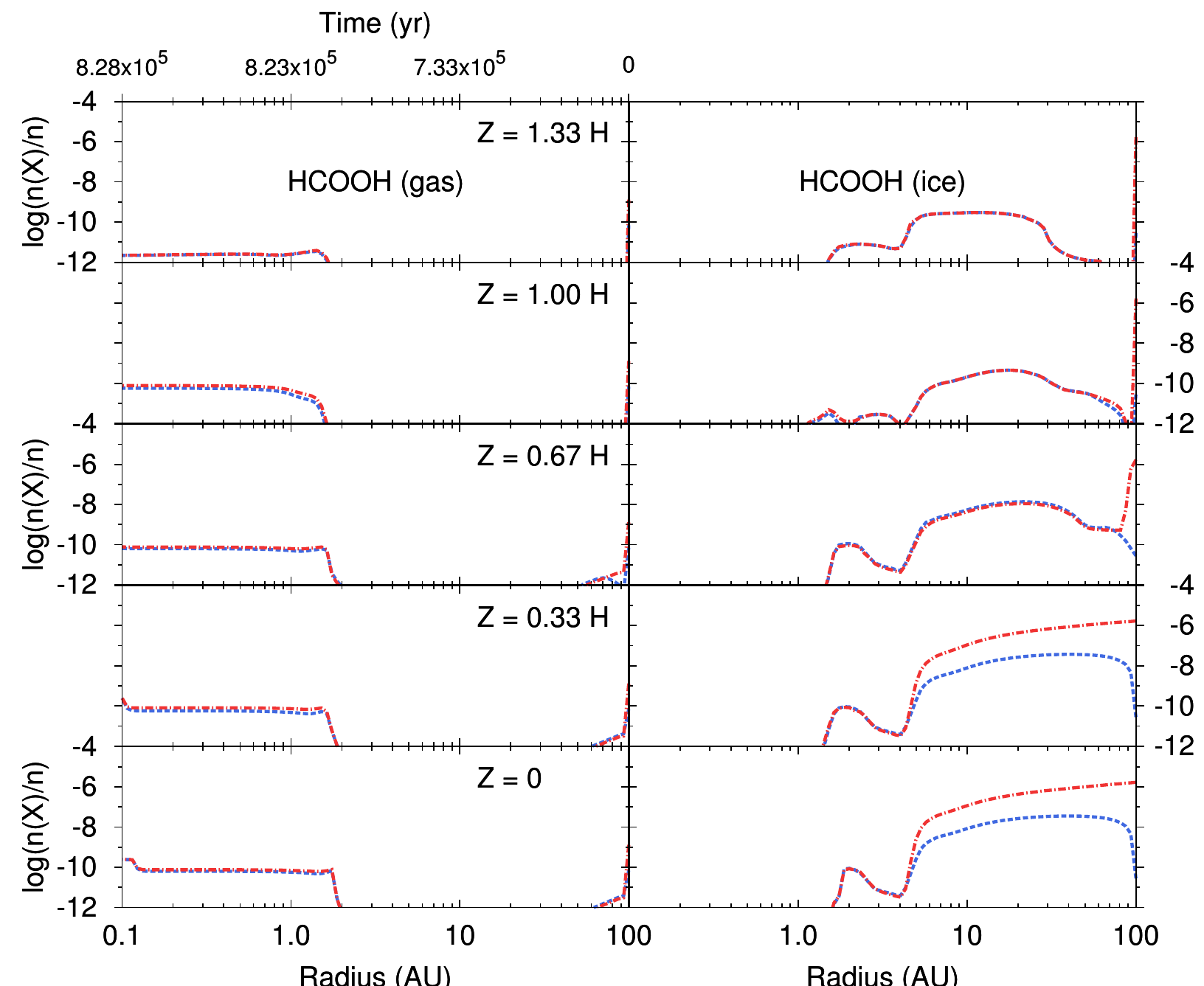}
\label{figure7}
\end{figure}

In Figures~\ref{figure6} and \ref{figure7} 
the fractional abundance of gas-phase (left-hand panel) and grain-surface (right-hand panel) 
formic acid, \ce{HCOOH}, is shown as a function of disk radius for the isolated and irradiated disk models, 
respectively.  
The story for formic acid is different from that for methanol.  
Formic acid is postulated to possess several routes to formation on grain surfaces at low 
temperatures, including the radical-radical association reaction, 
\begin{equation}
\ce{HCO} + \ce{OH} \longrightarrow \ce{HCOOH}.
\end{equation}
HCOOH has also been postulated to form via hydrogenation of the `HOCO' complex, \cite{ioppolo11} {\em i.e.},
\begin{align}
\ce{CO}   + \ce{OH} &\longrightarrow \ce{HOCO}, \\
\ce{HOCO} + \ce{H}  &\longrightarrow \ce{HCOOH}. 
\end{align}
\citeauthor{oberg09c} studied the formation of HCOOH via the first reaction during the 
irradiation of pure methanol ice. 
Upper limits only were determined due to the difficulty uniquely 
identifying HCOOH bands in the RAIRS (reflection-absorption infrared spectroscopy) 
spectrum. \cite{oberg09c}
\citeauthor{ioppolo11} later studied the formation of HCOOH via the 
second pathway and determined it was efficient at low temperatures 
$\lesssim$~20~K. \cite{ioppolo11} 
Both pathways are included in the network used here; however, 
the reaction, \ce{CO}~+~\ce{OH}~$\rightarrow$~\ce{HOCO}, 
has a large reaction barrier, $\gtrsim$~1500~K, originating 
from studies of the gas-phase reaction potential energy surface.\cite{chen05}
Because thermal grain-surface chemistry is used, 
this reaction is slow at low temperatures and HCOOH 
is not efficiently formed at $\sim$~10~K. 
As a consequence, the distribution of HCOOH along the accretion flow is sensitive 
to the assumed initial abundance.    

In the isolated disk, the abundance of HCOOH ice and gas for the warm case 
is around two orders of magnitude higher than for the cold case.  
A comparison with the initial abundances shows that grain-surface HCOOH is formed in the 
outer disk for the cold case in both models albeit reaching a 
lower abundance than that achieved for the warm case.  
In the isolated disk, as found for methanol, 
the formic acid fractional abundance, $\sim$~10$^{-7}$~--~10$^{-6}$, 
is preserved along the accretion flow.    
The snowline for HCOOH also resides at a similar radius to that for 
methanol ($\approx$~2~AU) reflecting their similar 
binding energies (see Table~\ref{table1}).     
At higher scale heights in the isolated disk, formic acid is photodesorbed 
reaching a gas-phase fractional abundance $\sim$~10$^{-9}$.  
Formic acid ice transported into the inner regions of isolated protoplanetary disks 
may have an interstellar origin provided the initial abundance entering the disk is 
sufficiently high.  
HCOOH ice has been observed in the envelopes of several low-mass protostars 
with abundances $\sim$~1~-~5\% of the water ice abundance.  \cite{boogert08}

In the irradiated disk, the gas-phase formic acid abundance in the inner 
region of the disk does not reflect that injected into the outer disk.  
The higher temperature ($>$~60~K) in the midplane helps thermally 
process the ice such that the cold and warm models converge 
at a radius of $\approx$~5~AU, i.e., outside the snowline at $\approx$~2~AU.  
This thermal processing leads to a drop in the fractional abundance of 
formic acid ice from $\sim$~10$^{-8}$~-10$^{-6}$ to 10$^{-10}$ at $\approx$~5~AU.  
Hence, the thermally desorbed formic acid inside of the snowline 
reaches a peak fractional abundance $\sim$~10$^{-10}$.    
Molecules which rely on radical-radical association reactions are 
sensitive to the higher temperature because radicals generally possess a lower binding energy 
to the grain surface than molecules, e.g., the binding energy assumed here for 
OH and HCO are 2850~K and 1600~K, respectively.  
Whether a molecule can form efficiently via this process relies on a sufficient 
supply of precursor radicals on the grain surface.  
At higher temperatures, the thermal desorption of radicals begins to 
compete with the thermal grain-surface association rates.  
For the case of formic acid in the irradiated disk midplane, 
this occurs at $\approx$~80~K (at $\approx$~5~AU).   
 
\subsubsection{Acetaldehyde.~~} 

\begin{figure}
\caption{Fractional abundance (with respect to number density) 
of \ce{CH3CHO} gas and ice along each streamline as a function of 
disk radius for the isolated disk model. 
The blue lines show results from the model using the `cold' (10~K) set of initial abundances 
and the red lines from the model using the `warm' (30~K) set of initial abundances (see, Table~\ref{table1}).}
\includegraphics[width=\textwidth]{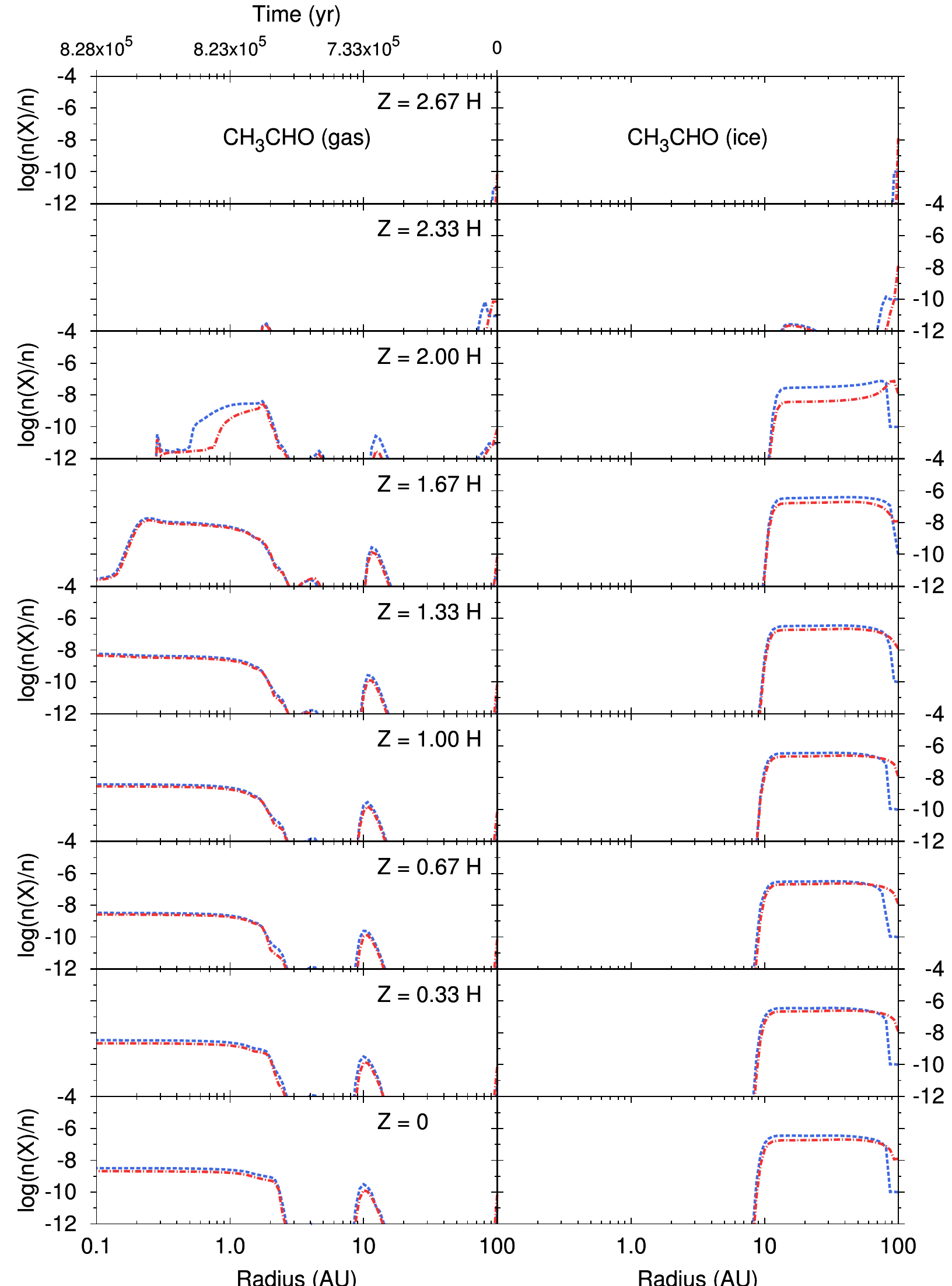}
\label{figure8}
\end{figure}

\begin{figure}
\caption{Fractional abundance (with respect to number density) 
of \ce{CH3CHO} gas and ice along each streamline as a function of 
disk radius for the irradiated disk model. The blue lines show results from the model using the `cold' (10~K) set of initial abundances 
and the red lines from the model using the `warm' (30~K) set of initial abundances (see, Table~\ref{table1}).}
\includegraphics[width=\textwidth]{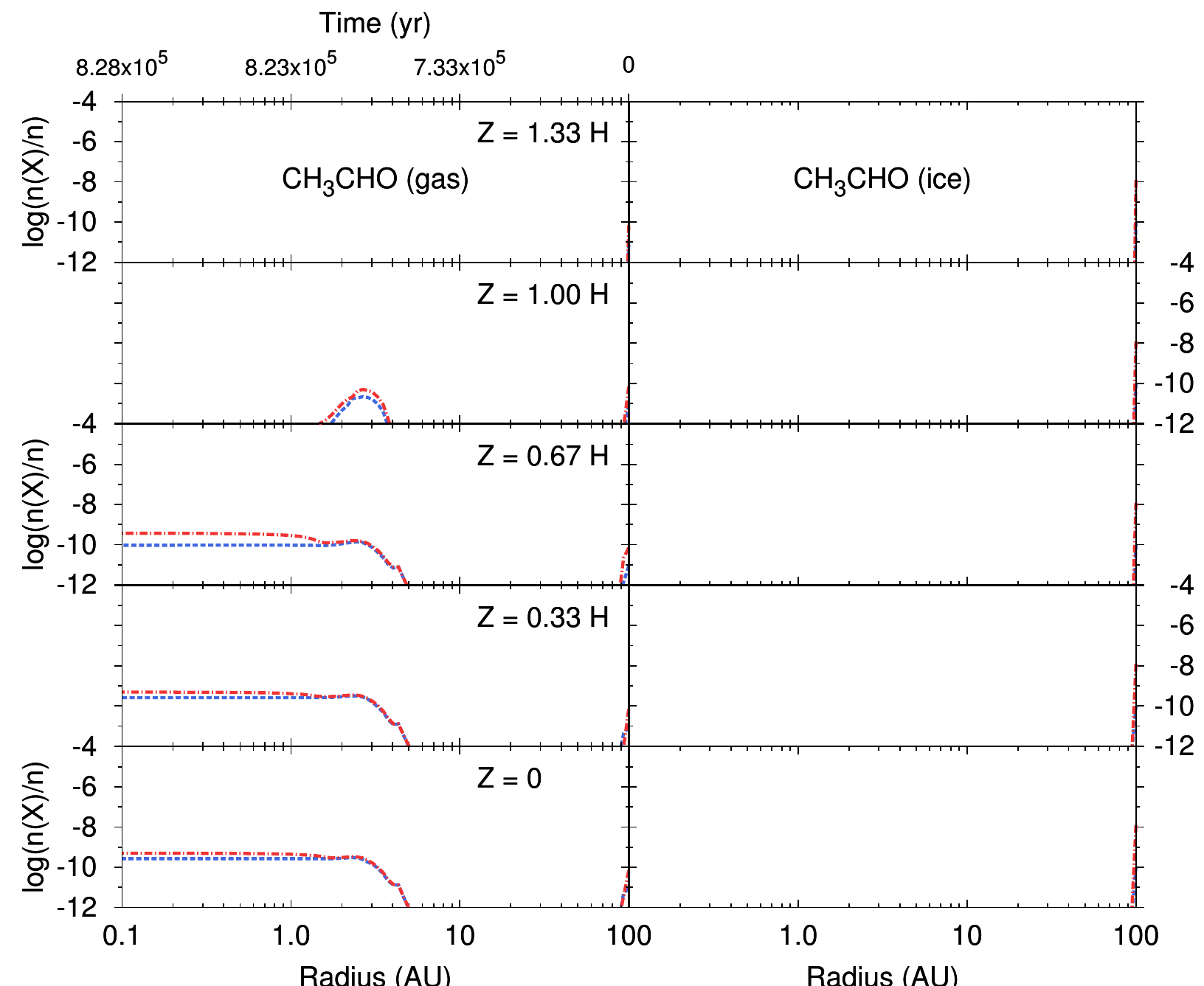}
\label{figure9}
\end{figure}

In Figures~\ref{figure8} and \ref{figure9} 
the fractional abundance of gas-phase (left-hand panel) and grain-surface (right-hand panel) 
acetaldehyde, \ce{CH3CHO}, is shown as a function of disk radius for the isolated and irradiated disk models, 
respectively.  

In contrast to formic acid, acetaldehyde is not sensitive to the assumed
initial abundances. 
In the isolated disk for the cold case, \ce{CH3CHO} forms efficiently 
in the outer disk midplane reaching a similar ice fractional abundance 
as for the warm case, $\sim$~10$^{-7}$, for radii $\gtrsim$~10~AU.  
Acetaldehyde ice can from on the grain via the association of the \ce{CH3} 
and \ce{HCO} radicals.  
This reaction is barrierless and so can proceed at the low temperatures 
in the outer regions of the isolated disk.  
The snowline for acetaldehyde resides at $\approx$~10~AU reflecting 
its lower binding energy (2780~K).  
There is a narrow thermally desorbed region of gas-phase acetaldehyde 
at $\approx$~10~AU.  
Within this radius, the temperature is sufficiently high for efficient 
thermal desorption of both the \ce{CH3} and \ce{HCO} radicals so that 
the radical-radical association formation pathway is no longer effective.  
Acetaldehyde can also form via hydrogenation of \ce{CH3CO} which itself 
forms via the association of \ce{CH3} and \ce{CO}; however, this reaction 
has a large reaction barrier ($\approx$~3500~K) and CO is also a volatile species 
with a binding energy 1150~K.  
The origin of gas-phase acetaldehyde in the inner disk midplane ($\lesssim$~2~AU) 
is thus gas-phase chemistry induced by the thermal desorption of 
strongly bound COMs, for example,  \ce{C2H5OH}.  
Gas-phase acetaldehyde forms primarily via dissociative electron 
recombination of the protonated form, \ce{CH3CH2O+}, which 
can form via reaction of cations with \ce{C2H5OH}.   

Acetaldehyde ice in the outer region of the disk 
may have an interstellar origin, again, if the injected abundance 
is sufficiently high.  
There are no reported detections of acetaldehyde ice in molecular 
clouds or protostellar envelopes.  
However, gas-phase acetaldehyde has been detected towards several 
dark clouds and prestellar cores with a fractional abundance 
$\sim$~10$^{-11}$~--~10$^{-10}$ 
with respect to the \ce{H2} gas density. \cite{oberg10a,bacmann12,cernicharo12}
The relatively constant abundance across sources suggests 
the source of \ce{CH3CHO} is grain-surface formation followed by 
non-thermal desorption into the gas phase. 

In the irradiated disk, the temperature is sufficiently high to 
thermally desorb \ce{CH3CHO} (and its precursor radicals) 
such that the molecule survives neither on the grain, nor in the gas.  
\ce{CH3CHO} molecules destroyed in the gas phase cannot be replenished 
via grain-surface chemistry due to the volatile nature of \ce{CH3} and \ce{HCO}.  
This result suggests \ce{CH3CHO} ice which is formed in comets 
within externally irradiated disks may have a secondary origin and is not pristine.  
The origin of gas-phase acetaldehyde in the inner disk is gas-phase 
formation following the thermal desorption of strongly bound COMs.  
Thus the fractional abundance of gas-phase acetaldehyde in the inner regions of 
the isolated disk and the irradiated disk are similar, $\sim$~10$^{-9}$.  

\subsubsection{Methyl formate.~~} 

\begin{figure}
\caption{Fractional abundance (with respect to number density) 
of \ce{HCOOCH3} gas and ice along each streamline as a function of 
disk radius for the isolated disk model. The blue lines show results from the model using the `cold' (10~K) set of initial abundances 
and the red lines from the model using the `warm' (30~K) set of initial abundances (see, Table~\ref{table1}).}
\includegraphics[width=\textwidth]{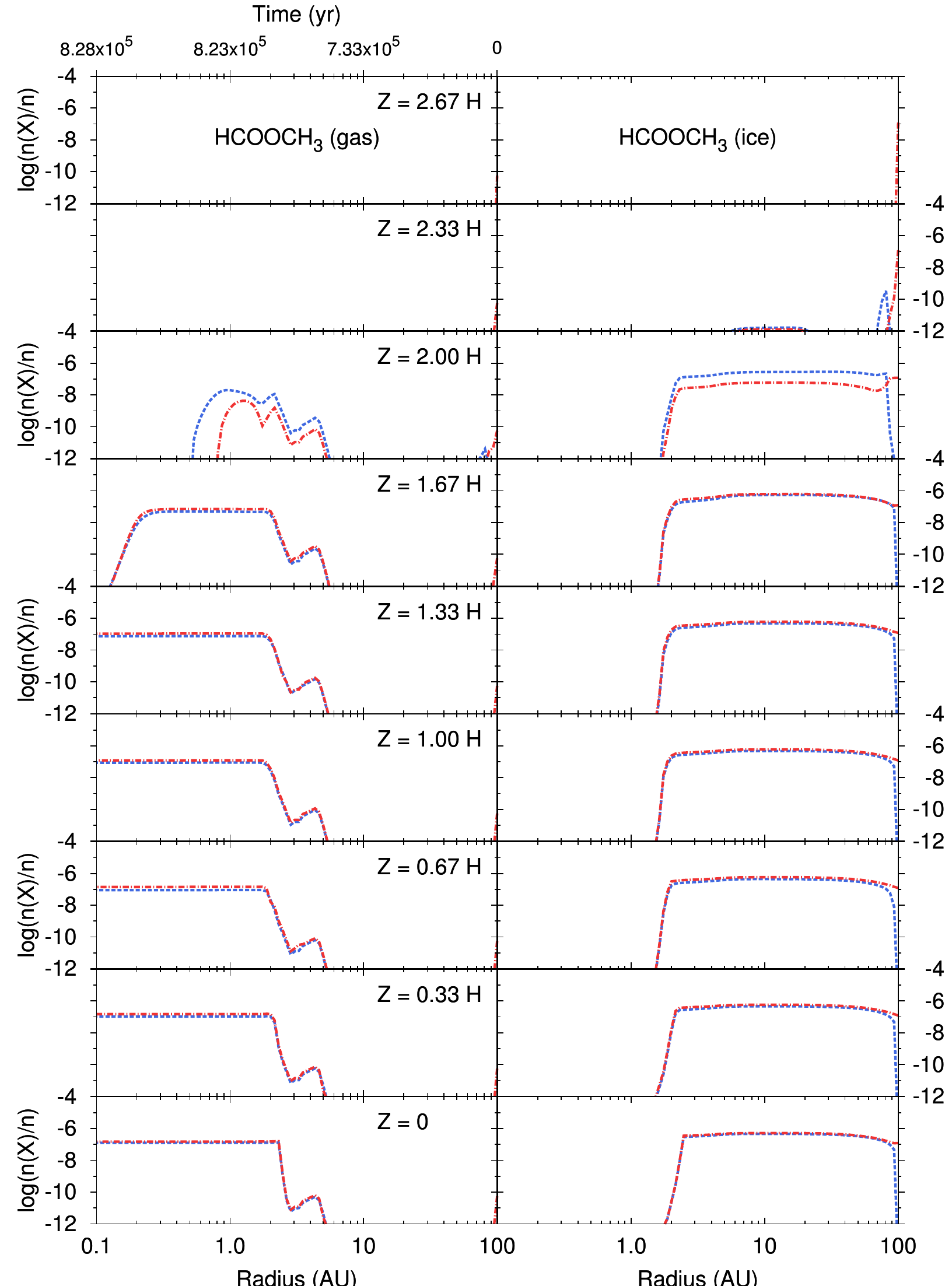}
\label{figure10}
\end{figure}

\begin{figure}
\caption{Fractional abundance (with respect to number density) 
of \ce{HCOOCH3} gas and ice along each streamline as a function of 
disk radius for the irradiated disk model. The blue lines show results from the model using the `cold' (10~K) set of initial abundances 
and the red lines from the model using the `warm' (30~K) set of initial abundances (see, Table~\ref{table1}).}
\includegraphics[width=\textwidth]{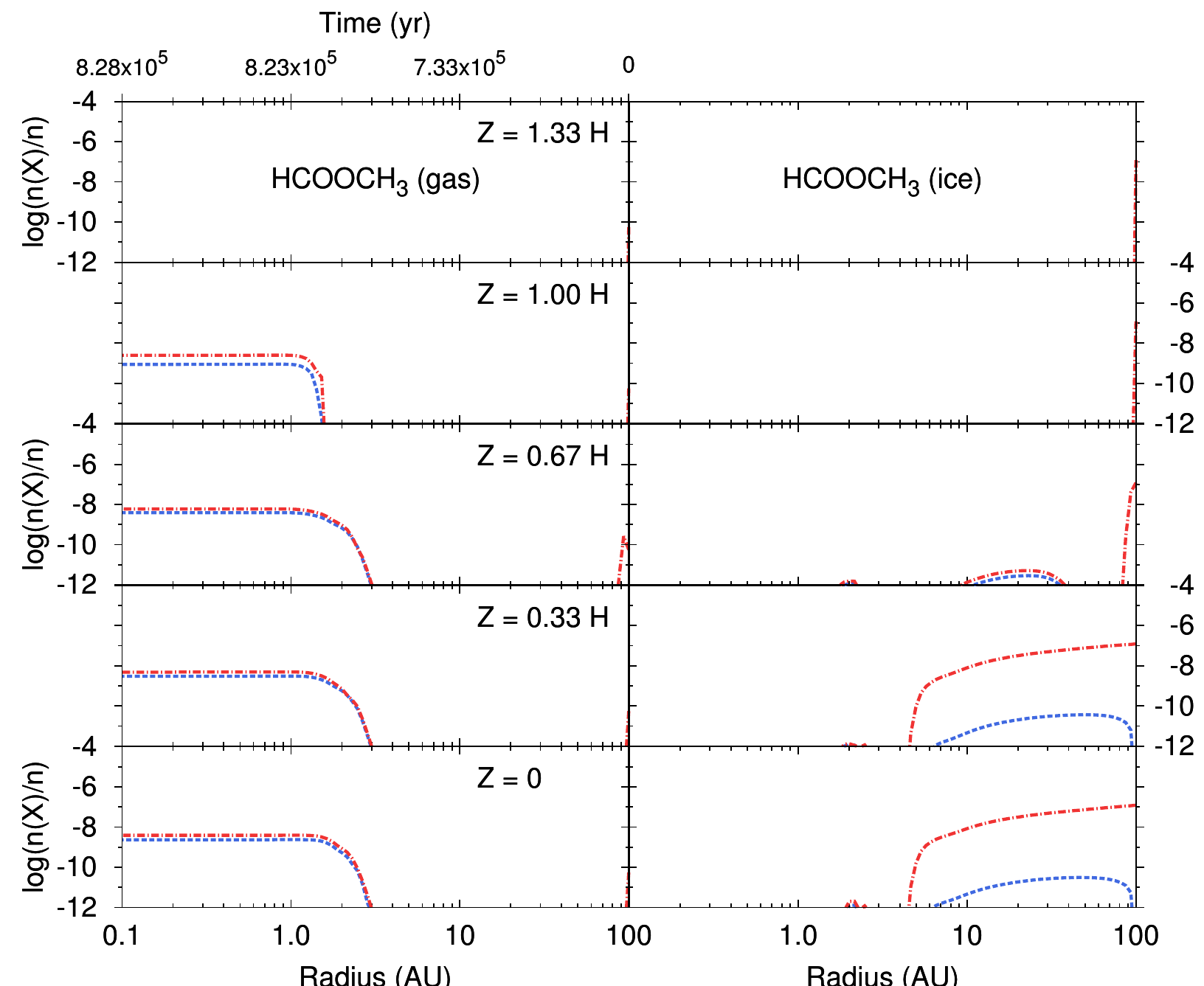}
\label{figure11}
\end{figure}

In Figures~\ref{figure10} and \ref{figure11} 
the fractional abundance of gas-phase (left-hand panel) and grain-surface (right-hand panel) 
methyl formate, \ce{HCOOCH3}, is shown as a function of disk radius for the isolated and 
irradiated disk models, respectively.  

For the isolated disk, the abundance of methyl formate is not sensitive 
to the assumed initial abundances: methyl formate is efficiently formed in the 
outer disk via the association of the \ce{CH3O} and \ce{HCO} radicals, reaching 
a peak fractional abundance, $\sim$~10$^{-7}$.    
This reaction is barrierless and can proceed at the temperatures in the 
outer disk, $\approx$~20~K.  
This agrees with the result found for the static model. \cite{walsh14}
The snowline for methyl formate lies slightly beyond the methanol snowline, between 
2 and 3~AU.  
This reflects the slightly lower binding energy assumed for methyl formate (4100~K).  
The abundance of methyl formate is preserved in the inner disk region.  
The origin of methyl formate in the planet- and comet-forming regions of 
isolated disks may have an interstellar origin depending on the initial abundance entering 
the disk.  
Gas-phase methyl formate has been observed in dark clouds and prestellar 
cores with a fractional abundance similar 
to that for acetaldehyde $\sim$~10$^{-11}$~-~10$^{-10}$.\cite{oberg10a,bacmann12,cernicharo12}  

In the irradiated disk model, the abundance of methyl formate ice is sensitive to 
the assumed initial abundances.  
For the cold case, methyl formate ice reaches a peak fractional abundance $\sim$~10$^{-10}$ 
contrasted with $\sim$~10$^{-7}$ for the warm case.  
Again, the higher temperature in the irradiated disk leads to efficient desorption 
of the necessary precursor radicals, \ce{CH3O} and \ce{HCO}, which have assumed binding 
energies, 2500~K and 1600~K, respectively.  
Methyl formate thermally desorbs at a radius $\approx$~5~AU whereby it is destroyed 
by gas-phase chemistry.  
Within $\approx$~2~AU, gas-phase methyl formate is replenished via gas-phase chemistry 
induced by the thermal desorption of strongly bound COMs.  
Methyl formate can form in the gas via the dissociative recombination of the protonated 
form of the molecule, \ce{HCOOCH4+}. 
This species, in turn, can form via the following ion-neutral reactions, \cite{laas11}
\begin{align}
\ce{CH3OH2+} + \ce{HCOOH} &\longrightarrow \ce{HCOOCH4+} + \ce{H2O}, \\
\ce{HCOOH2+} + \ce{CH3OH} &\longrightarrow \ce{HCOOCH4+} + \ce{H2O}.
\end{align}
Thus, the formation of gas-phase methyl formate is triggered by the thermal desorption 
of \ce{CH3OH} and \ce{HCOOH} ice within $\approx$~2~AU.  
The peak fractional abundance attained is $\sim$~10$^{-9}$ compared with $\sim$~10$^{-7}$ for the 
isolated case.  
This result correlates well with previous work which concluded gas-phase formation of 
methyl formate contributes, at most, to $\sim$~1\% of the total gas-phase abundance 
in hot cores.\cite{laas11}

The scenario described here for methyl formate, \ce{HCOOCH3}, is similar to 
that for dimethyl ether, \ce{CH3OCH3}.  

\subsubsection{Glycolaldehyde.~~} 

\begin{figure}
\caption{Fractional abundance (with respect to number density) 
of \ce{HOCH2CHO} gas and ice along each streamline as a function of 
disk radius for the isolated disk model. The blue lines show results from the model using the `cold' (10~K) set of initial abundances 
and the red lines from the model using the `warm' (30~K) set of initial abundances (see, Table~\ref{table1}).}
\includegraphics[width=\textwidth]{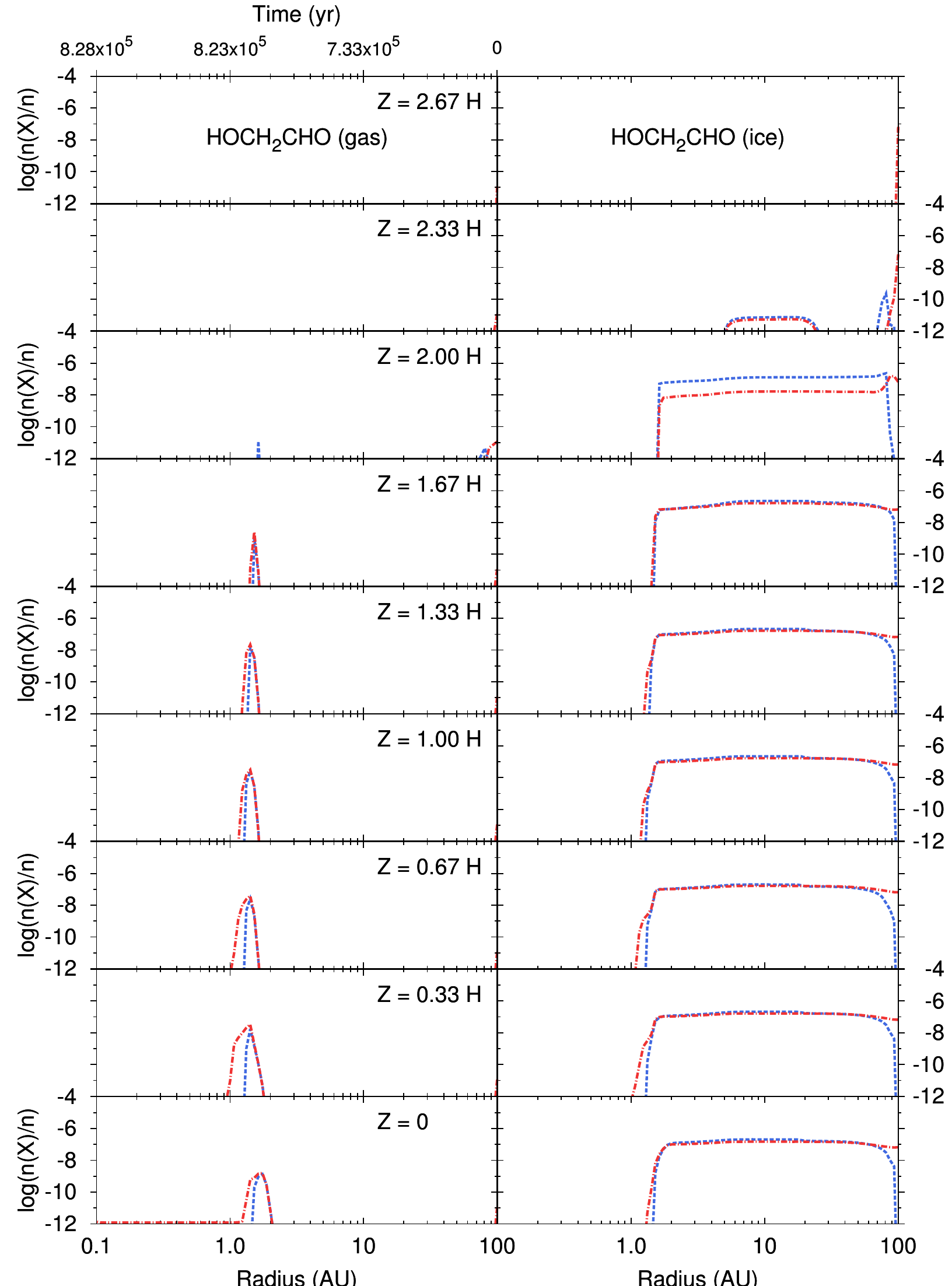}
\label{figure12}
\end{figure}

\begin{figure}
\caption{Fractional abundance (with respect to number density) 
of \ce{HOCH2CHO} gas and ice along each streamline as a function of 
disk radius for the irradiated disk model. The blue lines show results from the model using the `cold' (10~K) set of initial abundances 
and the red lines from the model using the `warm' (30~K) set of initial abundances (see, Table~\ref{table1}).}
\includegraphics[width=\textwidth]{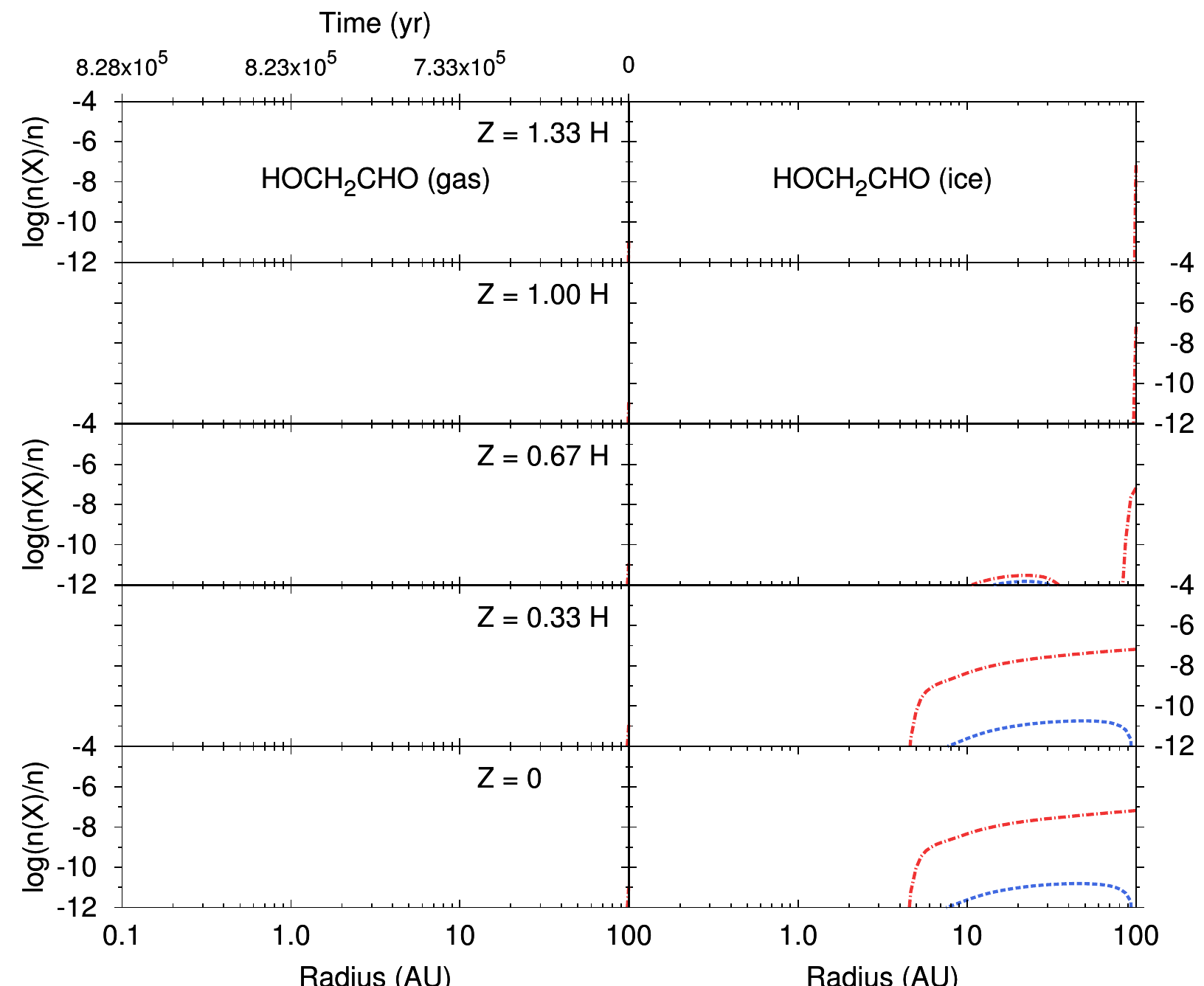}
\label{figure13}
\end{figure}

In Figures~\ref{figure12} and \ref{figure13} 
the fractional abundance of gas-phase (left-hand panel) and grain-surface (right-hand panel) 
glycolaldehyde, \ce{HOCH2CHO}, is shown as a function of disk radius for the 
isolated and irradiated disk models, respectively.  
Glycolaldehyde is an interesting complex molecule because it is the simplest 
molecule which possesses both an aldehyde and a hydroxyl group.  
It has also recently been detected in the gas phase towards the low-mass 
protostar, IRAS-16293 with a fractional abundance, $\sim$~10$^{-8}$ with 
respect to \ce{H2}. \cite{jorgensen12}

Similar to methyl formate, glycolaldehyde is not sensitive to assumed initial abundances 
in the outer regions of the isolated disk and can form efficiently via the barrierless reaction between 
\ce{CH2OH} and \ce{HCO}.  
Glycolaldehyde thermally desorbs within a radius of $\approx$~2~AU and exists in the gas phase in a narrow region 
between $\approx$~1 and 2 AU.  
Within this region, the conditions are such that grain-surface reformation 
cannot compete with gas-phase destruction on the timescales of the accretion flow; however, 
it is possible that the gas-phase chemical network is incomplete for 
larger COMs such as glycolaldehyde.  

The glycolaldehyde ice abundances in the outer region of the irradiated disk 
are very sensitive to the assumed initial abundances, again, requiring an 
appreciable abundance of precursor radicals on the grain.  
Once the glycolaldehyde desorbs, it is efficiently destroyed in the gas phase.    

\subsection{Comparison with previous disk models}

The influence of the accretion flow (or advection) on the gas-grain balance 
in protoplanetary disks 
has been investigated previously. \cite{aikawa99,ilgner04,nomura09,heinzeller11}
\citeauthor{aikawa99} computed the chemistry along 
streamlines from the outer disk ($\approx$~400~--~600~AU) to the inner disk  
($\approx$~10~--~70~AU) using a method similar to that outlined here.    
\citeauthor{ilgner04} and \citeauthor{heinzeller11} later 
investigated the chemical evolution of parcels of gas 
moving inwards along the accretion flow along multiple 
trajectories to build two-dimensional `snapshots' 
of the chemical structure of the planet-forming region of the disk ($\lesssim$~10~AU).  
In these works grain-surface chemistry was not included which 
restricted the discussion to small, simple species.    
\citeauthor{nomura09} considered the chemistry and transport of 
COMs in a protoplanetary disk for the first time.  
However, as in previous works, a grain-surface network was not 
explicitly included and the calculations followed the subsequent 
gas-phase chemistry upon thermal desorption of the parent COMs
assumed to be already present in the ice mantle.  
The discussion was also restricted to the inner planet-forming zone 
($\lesssim$~12~AU).  
Hence, the work presented here is the first investigation of 
grain-surface chemistry and the subsequent formation 
and destruction of COMs along the accretion flow in a protoplanetary disk.

Recently, the chemical evolution 
along parcels of gas in a hydrodynamical simulation of a gravitationally unstable 
embedded disk were reported for the first time.  \cite{ilee11} 
The parcels were found to follow erratic paths 
encountering a range of physical 
conditions and spikes in temperature and density due to shocks.  
The resulting molecular abundances trace density enhancements in spiral arms 
and a subsequent calculation of the molecular line emission suggested 
these structures may be observable with modern interferometric facilities 
such as ALMA.  \cite{douglas13} 
Again, grain-surface chemistry was not included in these calculations,  
and the influence on the abundance and distribution of COMs in 
hydrodynamic simulations of protoplanetary disks is yet to 
be investigated.

There are several additional physical 
processes thought to be important in disks which can influence the 
physical structure and resulting chemistry. \cite{henning13}
The effects of turbulent mixing on the composition of protoplanetary disks 
has been studied by numerous groups. 
\cite{ilgner04,semenov06,willacy06,aikawa07,heinzeller11,semenov11}
Early work suggested that vertical mixing does not change the 
chemically stratified nature of the disk; however, mixing 
can significantly increase the depth of the molecular layer 
and thus the vertical column densities. \cite{willacy06} 
Of these works, only one has included the formation of complex organic molecules 
via thermal grain-surface chemistry.  \cite{semenov11}
It was found that the abundances of COMs are very sensitive to 
the treatment of turbulent mixing, a similar conclusion to that 
presented here for the case of advection.  
A chemical model including both processes would be ideal; however, 
such models remain computationally challenging.  

The dust evolution in the disk 
can have a profound effect on the disk physical structure and 
resulting chemistry. 
Grain growth (coagulation) skews the dust-grain size distribution towards 
larger grains whereas settling (sedimentation) depletes the surface layers 
of large grains and increases the relative abundance of large grains 
in the disk midplane. \cite{dominik97,dullemond04}
Several groups have investigated the influence of dust-grain evolution on 
disk chemistry. \cite{jonkheid04,aikawa06,fogel11,vasyunin11,akimkin13}
The inclusion of grain growth tended to push the molecular layer deeper into the 
disk with increasing grain size; however, the vertical column densities 
of molecules remained unaffected. \cite{aikawa06}
Grain settling was found to lead to a smaller 
freeze out zone and a larger molecular layer due to the increased
penetration of UV radiation. \cite{fogel11} 
Recent models have coupled 
both dust evolution processes in a disk model with full gas-grain 
chemistry. \cite{vasyunin11,akimkin13}
In these models the chemically stratified nature 
of the disk is retained, albeit with a molecular layer which 
resides deeper into the disk.  
Both works also found that the gas-phase column densities are enhanced 
(at the expense of the grain-surface column densities) 
relative to the results using pristine interstellar dust.  
Grain coagulation may affect the formation rate of COMs due to the 
reduced surface area available for freeze out of the necessary precursor 
species.  
On the other hand, grain settling increases the dust-to-gas 
mass ratio in the disk midplane relative to that higher in the 
disk atmosphere, somewhat counteracting the reduction in total surface area.  
In addition, the deeper penetration of UV radiation may 
increase the efficiency of both thermal and non-thermal desorption 
of COMs formed on the grain into the gas phase. 
The combined effect of accretion flow and dust evolution 
on the abundance and distribution of COMs in protoplanetary disks should be 
explored in the future.

\section{Conclusions}
\label{conclusion}

The results presented here suggest that COMs which have efficiently formed 
on grain surfaces under cold, i.e., prestellar, conditions survive the transport 
along the accretion flow in the midplane of isolated disks; hence, the composition 
of icy planetesimals in isolated disks may be representative of the pristine 
interstellar ice entering the outer shielded regions of the disk, provided these 
species reach appreciable abundances under prestellar conditions.  
This conclusion is similar to that reached for simple ice species,  e.g., \ce{H2O}, 
in investigations on the influence of collapse, infall, and subsequent disk 
formation in low-mass protostars.\cite{visser09a}  Currently, a similar 
investigation is underway for more complex ice species.\cite{drozdovskaya14}
The successful identification of gas-phase COMs such as \ce{CH3OH}, \ce{CH3CHO}, \ce{HCOOCH3}, 
and \ce{CH3OCH3} in dark clouds and prestellar cores also suggest that grain-surface 
chemistry can operate effectively in dense, cold environments. \cite{oberg10a,bacmann12,cernicharo12}
However, it should be noted that, of the species considered in this work, 
only methanol ice and formic acid ice have been detected in prestellar environments.  
\cite{schutte99,gibb00,pontoppidan03,gibb04,boogert08,boogert11} 

If simple ices only are injected into the outer disk, the physical conditions 
facilitate the efficient production of COMs such as those considered here:   
\ce{CH3CHO}, \ce{HCOOCH3}, \ce{CH3OCH3}, and \ce{HOCH2CHO}.  
Hence, there is an increase in molecular complexity from cloud to disk, producing 
abundances of complex molecules similar to those observed in cometary comae.\cite{bockeleemorvan04}
This conclusion is similar to that reached in the static disk model.\cite{walsh14}
The chemical timescales are sufficiently short in the outer disk 
that the results from the static model and accretion model (presented here) 
are similar but not identical.  
The fractional abundances of grain-surface COMs 
along the disk midplane in the static model vary with radius: 
the long lifetime of the disk allows the chemistry 
to settle to the local physical conditions.  
Hence, the assumed initial abundance of COMs are not 
preserved in the static model as is seen in the accretion flow model
presented here. 
This work has shown that the grain-surface chemistry is sensitive 
to both the initial abundances adopted in the model and the accretion flow.

Once the snowline of each species is breached, the COMs can thermally desorb 
with the gas-phase abundance reflecting that injected from the ice mantle 
as also found in previous work investigating the transport of methanol ice in a disk.\cite{nomura09} 
This is in stark contrast to that found in the static model in which 
the extreme densities and temperatures encountered in the inner midplane are 
able to destroy gas-phase COMs over the lifetime of the disk $\sim$~10$^{6}$~years.
In this model, the gas spends only $\approx$~5000~years within 1~AU of the star.  
In order to properly treat the transport and chemistry of complex molecules 
within the planet-forming region of isolated protoplanetary disks, 
the influence of the accretion flow must be taken into account. 
In the disk molecular layer, COMs photodesorb into the gas phase, reaching 
peak fractional abundances similar to those found in the static model.\cite{walsh14}

Conversely, the radiation field in the irradiated disk is too strong for 
photodesorbed molecules to survive in the gas phase.  
It is also found that the assumed initial abundances are more important 
for this model due to the much higher temperatures in the outer disk.  
If simple ices are injected, thermal grain-surface chemistry is unable 
to form complex molecules because the necessary 
precursor radicals, which tend to be particularly volatile, can desorb efficiently.  
If more complex ices are injected into the outer disk, the grain-surface COMs
undergo thermal processing such that the gas phase abundances in the inner 
region do not reflect those injected at the outer edge of the disk.  
The origin of gas-phase COMs in the inner region of the irradiated disk 
is gas-phase formation induced by the thermal desorption of strongly bound 
molecules,such as \ce{CH3OH}, which do not rely on radical-radical formation.  
This could mean that icy planetesimals which form in the midplane of irradiated disks 
are likely to be more thermally processed and also composed of more simple 
ices than those in isolated disks.  
This is interesting given that the Sun has been postulated to have formed in 
a stellar cluster which has long since dispersed.\cite{adams10}  
If the comets in the Solar System were originally composed of more simple ices, their 
current composition could reflect post-formation processing of the cometary surface.\cite{mumma11}

The work presented here does not address one of the outstanding issues 
in grain-surface chemistry, that is, the efficient 
formation at 10~K of those complex molecules which are thought to form via 
radical-radical association reactions.  
Under the current paradigm, molecular radicals do not have sufficient 
mobility to diffuse within or on the ice mantle.  
There are several processes which have been postulated to help 
build chemical complexity at 10~K including the radiation-processing of ices via 
UV photons and cosmic rays,\cite{oberg09c,bennett07} and  
a high efficiency of reactive desorption.\cite{vasyunin13,dulieu13}
These issues will be explored, in the context of protoplanetary disk chemistry, in future work.  

\section*{Acknowledgements}
C.W. acknowledges support from the European Union A-ERC grant 291141
CHEMPLAN and financial support (program number 639.041.335)
from the Netherlands Organisation for Scientific Research (NWO). 
EH wishes to acknowledge the support
of the National Science Foundation for his astrochemistry program, and his program 
in chemical kinetics through the Center for the
Chemistry of the Universe. He also acknowledges support from the NASA Exobiology and 
Evolutionary Biology program through a subcontract from Rensselaer Polytechnic Institute.
H.~N.~acknowledges the Grant-in-Aid for Scientific Research 23103005 and 25400229.  
She also acknowledges support from the Astrobiology Project of the CNSI, NINS (Grant Number 
AB251002, AB251012). 
Astrophysics at QUB is supported by a grant from the STFC.  
S.L.W.W. acknowledges support from startup funds provided by Emory University.

\footnotesize{
\bibliography{faraday_arxiv}
\bibliographystyle{rsc}
}

\end{document}